\documentclass[onecolumn]{article}%
\usepackage{amsmath}
\usepackage{amsfonts}
\usepackage{amssymb}
\usepackage{graphicx}
\usepackage{revsymb}%
\begin{document}

\title{Spin Foams for the Real, Complex Orthogonal Groups in 4D and the bivector
scalar product reality constraint.}
\author{Suresh. K.\ Maran}
\maketitle

\begin{abstract}
The Barrett-Crane model for the $SO(4,C)\ $general relativity is
systematically derived. This procedure makes rigorous the calculation of the
Barrett-Crane intertwiners from the Barrett-Crane constraints of both real and
complex Riemannian general relativity. The reality of the scalar products of
the complex bivectors associated with the triangles of a flat four simplex is
equivalent to the reality of the associated flat geometry. Spin foam models in
4D for the real and complex orthogonal gauge groups are discussed in a unified
manner from the point of view of the bivector scalar product reality
constraints. Many relevant issues are discussed and generalizations of the
ideas are introduced. The asymptotic limit of the $SO(4,C)$ general relativity
is discussed. The asymptotic limit is controlled by the $SO(4,C)$ Regge
calculus which unifies the Regge calculus theories for all the real general
relativity cases. The spin network functionals for the 3+1 formulation of the
spin foams are discussed. The field theory over group formulation for the
Barrett-Crane models is discussed briefly. I introduce the idea of a mixed
Lorentzian Barrett-Crane model which mixes the intertwiners for the Lorentzian
Barrett-Crane models. A mixed propagator is calculated. I also introduce a
multi-signature spin foam model for real general relativity which is made by
splicing together the four simplex amplitudes for the various signatures is
defined. Further research that is to be done is listed and discussed.

\end{abstract}
\tableofcontents

\section{Introduction}

The idea of spin foams \cite{JCBintro}, \cite{PerezReview} briefly reviewed in
this section is a proposal for background independent, non-perturbative and
coordinate independent quantum general relativity\footnote{I refer the novice
readers to the latest book by Rovelli \cite{CRBook} on background independent
quantum gravity for startup.}. Spin foams are essentially the path integral
quantizations of general relativity and related theories on simplicial
manifolds \cite{JCBintro}, \cite{ooguriBFderv}, \cite{FoamDer},
\cite{ClassActFoam}. Spin foams have various advantages. They are simply
combinatorics and so they do not require a back-ground space-time to exist.
Otherwise, a spin foam is a `thing-in-itself' and so an ultimate object in
terms of which reality could be understood. Spin foam models are connected to
classical general relativity through Regge calculus \cite{ReggeCalc}. Spin
foam quantization is similar to lattice gauge theory. The later has been found
useful in many issues in QCD \cite{QCD}. A variation of Regge calculus called
the dynamical triangulation has shown promising semiclassical limit
\cite{Lollfirstprincples}.

\subsection{Review of Spin Foams}

Spin foams are dynamical generalization of the idea of spin networks
\cite{Pen} to higher dimensional space \cite{JCBintro}. The essence of spin
networks and spin foams is gauge invariance. An abstract closed spin foam in
$N$ dimensions for a group $G$ is based on the following
constructions\footnote{Please refer to Baez \cite{JCBintro} for a more
technical definition.} (in simple terms):

\begin{itemize}
\item Consider a $N$ dimensional closed oriented manifold triangulated using
$N$ dimensional simplices.

\item With each $(N-2)$-simplex $b$ associate an irreducible representation
$\rho_{b}$ of the group $G$.

\item With each $(N-1)$-simplex $e$ associate an intertwiner $i_{e}$ of the
group that intertwines the representations associated with its $N-2$ simplices.
\end{itemize}

Spin foams are usually defined using a partition function. A typical
definition of a spin foam partition function is%

\[
Z(\Delta)=\sum\limits_{\{\rho_{b},i_{e}\}}A_{N-2}(\rho_{b})A_{N-1}(\rho
_{b},i_{e})A_{N}(\rho_{b},i_{e}),
\]
where the $\Delta$ denotes a triangulation. The sum is over all possible
intertwiners and representations associated with the simplices. The $A$'s are
the quantum amplitudes associated with the simplices of the various
dimensions. The $A_{N}(\rho_{b},i_{e})$ is the amplitude associated with a
$N$-simplex as a function of the intertwiners and the representations
associated with its lower dimensional simplices. Usually this amplitude is
given by a spin network built using the intertwiners, the representations and
the dual graph to the triangulation of the $N$-simplex. The $A_{N-1}(\rho
_{b},i_{e})$ is the amplitude of a $(N-1)$-simplex as a function of the
intertwiner associated with it and the representations associated with its
$(N-2)$-simplices. The $A_{N-2}(\rho_{b})$ is the amplitude of a
$(N-2)$-simplex as a function of the representation associated with it.

The physics related to a spin foam is contained in the definition of its
amplitudes. There are various possible spin foam models available based on the
various definitions for the amplitudes \cite{ClassActFoam}, \cite{PerezReview}%
. The amplitudes associated with these models can be derived by the path
integral quantizations of discretized actions \cite{JCBintro}, \cite{FoamDer},
\cite{ClassActFoam}, \cite{ooguriBFderv}.

An important class of spin foam models are those of the topological field
theory called the BF theory\footnote{A BF theory in $n$ dimensions and for a
group $G$ refers to a field theory defined by the action $S=\int B\wedge F$ .
Here the $B$ is a $n-2$ form which takes values in the dual Lie algebra of
$G$. The $F$ is a 2-form is the Cartan curvature of a $G$-connection $A$. The
free variables of the theory are the $B$ and $A$.} \cite{BF}. A spin foam
model for the four dimensional $SO(4,R)$ BF theory was derived directly from
its discretized action on simplicial manifolds by Ooguri \cite{ooguriBFderv}.
The quantization of the discrete Riemannian BF\ theory results in a partition
function which does not depend on the discretization of the four manifold
\cite{SpinFoamDiag}. This quantum model has only global degrees of freedom
like the classical BF theory \cite{JCBintro}.

In case of the four dimensional Riemannian general relativity Barrett and
Crane \cite{BCReimmanion} proposed a systematic way to assign an amplitude to
the four-simplex. They proposed a set of quantum constraints based on the
classical properties of the bivectors associated with the triangles of a flat
four-simplex. They also proposed a solution for these constraints
\cite{BCReimmanion}. The spin foam models constructed using the Barrett-Crane
procedure are called the Barrett-Crane models.

The Barrett-Crane model can be considered as the quantization of discretized
Plebanski formalism of general relativity \cite{Plebanski} on a simplicial
manifold. The Plebanski theory of general relativity is simply a four
dimensional BF theory combined with certain constraint called the Plebanski
constraint. This constraint enforces the $B$ field to be a wedge product of a
co-tetrad field with itself \cite{Plebanski}. The co-tetrad field contains the
metric information \cite{Plebanski}. The Barrett-Crane constraints contain the
information about the Plebanski constraint.

The Riemannian Barrett-Crane model can be formally derived starting from a
discretized action on a simplicial four manifold \cite{FoamDer}. It can be
obtained by deriving the Ooguri model and imposing the Barrett-Crane
constraints on it. Imposition of the Barrett-Crane constraints breaks the
topological nature of the Ooguri model and the discretization independence of
the theory. So the theory now acquires local excitations \cite{PerezReview}.

It is possible to rewrite the Riemannian Barrett-Crane four-simplex amplitude
of a four-simplex in terms of certain propagators on the homogenous
space\ $S^{3}=$ $SO(4)/SO(3)$. Spin foam model of the Lorentzian general
relativity were proposed by Barrett and Crane \cite{BCLorentzian}. These
models were constructed based on certain propagators on the homogenous spaces
of the Lorentz group corresponding to the various subgroups of it in the
Minkowski space, viz.

\begin{itemize}
\item The upper sheet of the double sheet hyperboloid: $H^{+}\approx
SL(2,\boldsymbol{C})/SU(2)$

\item The single sheet hyperboloid with the antipodal points considered as a
single point: $H^{-}\approx$ $SL(2,\boldsymbol{C})/U^{(-)}$ where
$U^{(-)}=SU(1,1)\otimes Z_{2}$.

\item The upper sheet of the null cone: $N$ $=$ $SL(2,\boldsymbol{C})/E(2)$
\end{itemize}

Rovelli and Perez proposed a way of deriving the first two models using the
field theory over group formulation \cite{RovPerGFTLorentz}, \cite{RoPeModel}.

\subsection{Motivation for this article}

There are two sets of issues at hand. The first set of issues relate to the
construction of spin foam models starting from general physical and
mathematical premises. Some of the issues involved here are:

\begin{enumerate}
\item How to understand the different spin foam models of general relativity
from a general point of view?

\item Are there other models that exist for Lorentzian general relativity?

\item Even though the Barrett-Crane constraints appear to have solutions, it
is not clear how to directly impose one of the constraints called the
cross-simplicity constraint\footnote{It is known that the cross-simplicity
condition implies that the internal representations of the Barrett-Crane
intertwiners must be simple \cite{BCReimmanion}. But the difficult part is the
simultaneous imposition of all the cross-simplicity constraints on a general
intertwiner in four dimensions.}.

\item The uniqueness of the Barrett-Crane model for the Riemannian general
relativity has been argued by Reisenberger \cite{ReisenBCinter}. How to do
this for the case of Lorentzian general relativity?

\item Is it possible to develop a unified understanding of the Barrett-Crane
models for the various signatures and the $SO(4,C)$ general relativity in four dimensions?

\item How to relate the ideas in the spin foams to canonical quantum general
relativity and vice versa. For example, whether the reality condition of
canonical quantum general relativity has any interpretation in spin foams?
\end{enumerate}

The second set of issues is about how to extract physics from spin foam
models. The two sets of issues are interlinked. This article is motivated by
the first set of issues. The second set of issues is discussed as future
research directions at the end of this article.

An attempt by me to rigorously develop and unify the various models for the
Lorentzian general relativity was made in Ref:\cite{MyRigorousSpinFoam}. The
attempt was made to derive the two models by directly solving the
Barrett-Crane constraints. The Barrett-Crane cross-simplicity constraint
operator was explicitly written using the Gelfand-Naimarck representation
theory. But after numerous attempts I\ could not obtain any solution for the
constraint. But the efforts in this research lead to the idea of the reality
for spin foam models. It also led to the systematic quantization method for
the Barrett Crane models of complex and real Riemannian general relativity.

Let us consider the Lorentzian Barrett-Crane models now. The Hilbert space of
the unitary representations of the Lorentz group $SL(2,C)$ is infinite
dimensional \cite{IMG}. A unitary representation of $SL(2,C)$ is labeled by a
complex number $\chi=$ $\frac{n}{2}+i\rho$ where $\rho$ is a real number and
$n$ an integer. The idea of simplicity requires $\rho n=0$ \cite{BCLorentzian}%
. Thus we are allowed to assign only one of either $\chi=\rho$ or $\chi
=i\frac{n}{2}$ to each triangles. Now consider the eigen-values of the Casimir
of $SL(2,C)$ in the complex form \cite{IMG},%
\[
\chi^{2}-1=-\rho^{2}+\frac{n}{4}^{2}+i\rho n.
\]
The $\rho n$ is precisely the imaginary part of the Casimir. So if $\chi
^{2}-1$ is interpreted as the square of the area of a triangle, then $\rho
n=0$ simply constrains the square of the area to be real. The situation is
further clarified if I start from the $SO(4,C)$ general relativity theory as
will be explained below.

The $SO(4,C)\ $Barrett-Crane model need to be constructed based on unitary
representation theory of the group $SO(4,C)$. The unitary representations of
$SO(4,C)$ can be constructed using the relation%
\begin{equation}
SO(4,C)\approx\frac{SL(2,C)\times SL(2,C)}{Z^{2}}. \label{GroupIso}%
\end{equation}
This is the complex analog of
\[
SO(4,R)\approx\frac{SU(2,C)\times SU(2,C)}{Z^{2}}.
\]

So similar to the unitary representation theory of the Riemannian group, the
unitary representations of $SO(4,C)$ can be labeled by two `$\chi$'s:
$(\chi_{L}=\rho_{L}+i\frac{n_{L}}{2},\chi_{R}=\rho_{R}+i\frac{n_{R}}{2}),$
where each $\chi$ represents a unitary representation of $SL(2,C)$ \cite{IMG},
$n_{L}+n_{R}$ even number (Please see appendix B for details).

The $SO(4,C)$ Barrett-Crane simplicity constraint sets one of the
$SO(4,C)$\ Casimir's eigen values $\left(  \chi_{L}^{2}-\chi_{R}^{2}\right)
/2=0$, which in turn sets $\chi_{L}=\pm\chi_{R}$ (=$\chi$ say). Then the other
Casimir's eigen value is
\[
\left(  \chi_{L}^{2}+\chi_{R}^{2}-2\right)  /2=\chi^{2}-1,
\]
which corresponds to the square of area. By setting this eigenvalue to be
real, we deduce the area quantum number that is to be assigned to a triangle
of a Lorentzian spin foam. So from the point view of the $SO(4,C)$
Barrett-Crane model the simplicity condition of the Lorentzian general
relativity is simply a reality condition.

The reality of the squares of areas can be imposed at the continuum classical
level by imposing the condition that the area metric to be real. Since the
area metric can be expressed as a function of a bivector field, this reality
constraint can be naturally combined with the Plebanski theory for the
$SO(4,C)$ general relativity. I have done this analysis in Ref:
\cite{ClassicalAreaReal}. There, I\ have shown that the area metric metric
reality condition reduces a complex metric to a real or imaginary metric. An
imaginary metric essentially describes a real geometry. I\ also have shown
there that one can derive real general relativity by adding a Lagrange
multiplier to the $SO(4,C)$ Plebanski action to impose the area metric reality constraint.

The idea of a Barrett-Crane intertwiner can be easily formalized. Then as will
be discussed in this article the models for real general relativity theories
for all signatures are related to that of the $SO(4,C)$ general relativity
through the quantum version of the discretized area metric reality
condition\footnote{The Barrett-Crane model based on the propagators on the
null-cone \cite{BCLorentzian} is an exception to this.}. In this way we have a
unified understanding of the Barrett-Crane models for the four dimensional
real general relativity theories for all signatures (non-degenerate) and the
$SO(4,C)$ general relativity. The discrete equivalent of the area metric
reality condition in the context of Barrett-Crane theory is that the scalar
products of bivectors associated with the triangles of a four- or
three-simplex be real \cite{ClassicalAreaReal}.

One of the new elements in the systematic derivation the $SO(4,C)$
Barrett-Crane model in this article is the rigorous imposition of the
Barrett-Crane cross-simplicity constraint on the intertwiners initially
defined as a function of many variables on the complex three sphere. This
procedure is directly applicable to the Riemannian Barrett-Crane model. Also
I\ calculate the asymptotic limit of the $SO(4,C)$ Barrett-Crane models and
extract the bivectors that satisfy all the Barrett-Crane Constraints excluding
the non-deneracy conditions.

In this article I\ discuss many ideas relating to the spin foams of the
$SO(4,C)$ and real general relativity listed in the layout below. \textbf{This
article makes rigorous, unifies and generalizes the Barrett-Crane spin foam
models of general relativity.}

\subsection{Article Layout}

\begin{itemize}
\item Section One: I discuss the spin foam model for the $SO(4,C)$ BF theory
based on Ooguri's research \cite{ooguriBFderv}.

\item Section two: I briefly discuss the continuum $SO(4,C)$ model
\cite{ClassicalAreaReal}. I\ call all the Barrett-Crane constraints excluding
the non-degeneracy conditions as the essential conditions. I\ call the
Barrett-Crane models obtained by quantizing these conditions as the essential
Barrett-Crane models. I develop the essential $SO(4,C)$ Barrett-Crane model by
solving the corresponding essential Barrett-Crane constraints. I\ explicitly
solve the Barrett-Crane cross-simplicity constraint on the function. I
describe the various properties of the propagators.

\item In section three using the bivector scalar product reality constraint
the Barrett-Crane models for the real general relativity for all signatures
and $SO(4,C)$ general relativity are discussed in a unified manner.

\item In section four I discuss various further developments.

\begin{itemize}
\item I discuss the asymptotic limit \cite{PonzanoReggeModel} of the $SO(4,C)$
Barrett-Crane model..

\item To relate the canonical quantum general relativity to the spin foams I
developed a $(N-1)+1$ model of the spin foams \cite{SMSKn-1+1}. In this model
the quantum partitions for general relativity and the BF\ theory can be
written down as the sum over amplitudes for histories of spin networks
functionals. This theory can be formally applied to various Barrett-Crane models.

\item I\ also briefly discuss the field theory \cite{GFTfirst},
\cite{GFTreimm} over group version of the $SO(4,C)$ Barrett-Crane model.

\item I introduce two possible new quantum real general relativity models. One
of them is a Lorentzian Barrett-Crane model and the other one is a
multi-signature real spin foam model.
\end{itemize}

\item In section five I briefly list the various new results in this article.
I also list and discuss the possible future works need to be done.
\end{itemize}

\section{Spin foam of the $SO(4,C)$ BF model}

Consider a four dimensional submanifold $M$. Let $A$ be a $SO(4,C)$ connection
1-form and $B^{ij}$ a complex bivector valued $2$-form on $M$. Let $F$ be the
curvature 2-form of the connection $A$. Then I define a real continuum
BF\ theory action,%
\begin{equation}
S_{BF}(A,B_{ij},\bar{A},\bar{B}_{ij})=\operatorname{Re}\int_{M}B\wedge F,
\label{BFaction}%
\end{equation}
where $A,B_{ij}$ and their complex conjugates are considered as independent
free variables. This classical theory is a topological field theory. This
property also holds on spin foam quantization as will be discussed below.

The Spin foam model for the $SO(4,C)$ BF\ theory action can be derived
from\ the discretized BF action by using the path integral quantization as
illustrated in Ref:\cite{ooguriBFderv} for compact groups. Let $\Delta$ be a
simplicial manifold obtained by a triangulation of $M$. Let $g_{e}\in SO(4,C)$
be the parallel propagators associated with the edges (three-simplices)
representing the discretized connection. Let $H_{b}=%
{\textstyle\prod_{e\supset b}}
g_{e}$ be the holonomies around the bones (two-simplices) in the four
dimensional matrix representation of $SO(4,C)$ representing the curvature. Let
$B_{b}$ be the $4\times4$ antisymmetric complex matrices corresponding to the
dual Lie algebra of $SO(4,C)$ corresponding to the discrete analog of the $B$
field. Then the discrete BF action is
\[
S_{d}=\operatorname{Re}\sum_{b\in M}tr(B_{b}\ln H_{b}),
\]
which is considered as a function of the $B_{b}$'s and $g_{e}$'s. Here $B_{b}$
the discrete analog of the $B$ field are $4\times4$ antisymmetric complex
matrices corresponding to dual Lie algebra of $SO(4,C)$. The $\ln$ maps from
the group space to the Lie algebra space. The trace is taken over the Lie
algebra indices. Then the\ quantum partition function can be calculated using
the path integral formulation as,%

\[
Z_{BF}(\Delta)=\int\prod_{b}dB_{b}d\bar{B}_{b}\exp(iS_{d})\prod_{e}dg_{e}%
\]%
\begin{equation}
=\int\prod_{b}\delta(H_{b})\prod_{e}dg_{e}, \label{eq.del}%
\end{equation}
where $dg_{e}$ is the invariant measure on the group $SO(4,C)$. The invariant
measure can be defined as the product of the bi-invariant measures on the left
and the right $SL(2,C)$ matrix components. Please see appendix A and B\ for
more details. Similar to the integral measure on the $B$'s an explicit
expression for the $dg_{e}$ involves product of conjugate measures of complex coordinates.

Now consider the identity
\begin{equation}
\delta(g)=\frac{1}{64\pi^{8}}\int d_{\omega}tr(T_{\omega}(g))d\omega,
\label{eq.del.exp}%
\end{equation}
where the $T_{\omega}(g)$ is a unitary representation of $SO(4,C),$ where
$\omega=$ $\left(  \chi_{L},\chi_{R}\right)  $ such that $n_{L}+n_{R}$ is
even, $d_{\omega}=\left\vert \chi_{L}\chi_{R}\right\vert ^{2}$. The details of
the representation theory is discussed in appendix B. The integration with
respect to $d\omega$ in the above equation is interpreted as the summation
over the discrete $n$'s and the integration over the continuous $\rho$'s.

By substituting the harmonic expansion for $\delta(g)$ into the equation
(\ref{eq.del}) we can derive the spin foam partition of the $SO(4,C)\ BF$
theory as explained in Ref:\cite{JCBintro} or Ref:\cite{ooguriBFderv}. The
partition function is defined using the $SO(4,C)$ intertwiners and the
$\{15\omega\}$ symbols.

The relevant intertwiner for the $BF\ $spin foam is%
\[
i_{e}=%
\raisebox{-0.3814in}{\includegraphics[
height=0.7628in,
width=0.518in
]%
{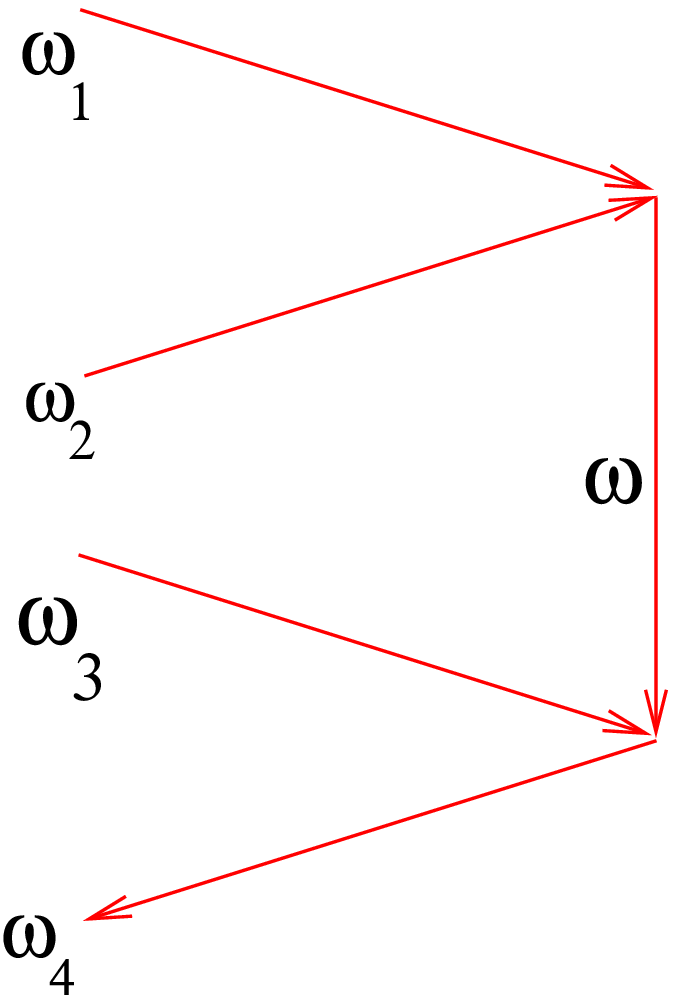}%
}%
.
\]
The nodes where the three links meet are the Clebsch-Gordan coefficients of
$SO(4,C)$. The Clebsch-Gordan coefficients of $SO(4,C)$ are just the product
of the Clebsch-Gordan coefficients of the left and the right handed $SL(2,C)$
components. The Clebsch-Gordan coefficients of $SL(2,C)$ are discussed in the
references \cite{IMG} and \cite{NaimarckClebsch}.

The quantum amplitude associated with each simplex $s$ is given below and can
be referred to as the $\{15\omega\}$ symbol,%

\[
\{15\omega\}=%
\raisebox{-0.8026in}{\includegraphics[
height=1.548in,
width=1.6016in
]%
{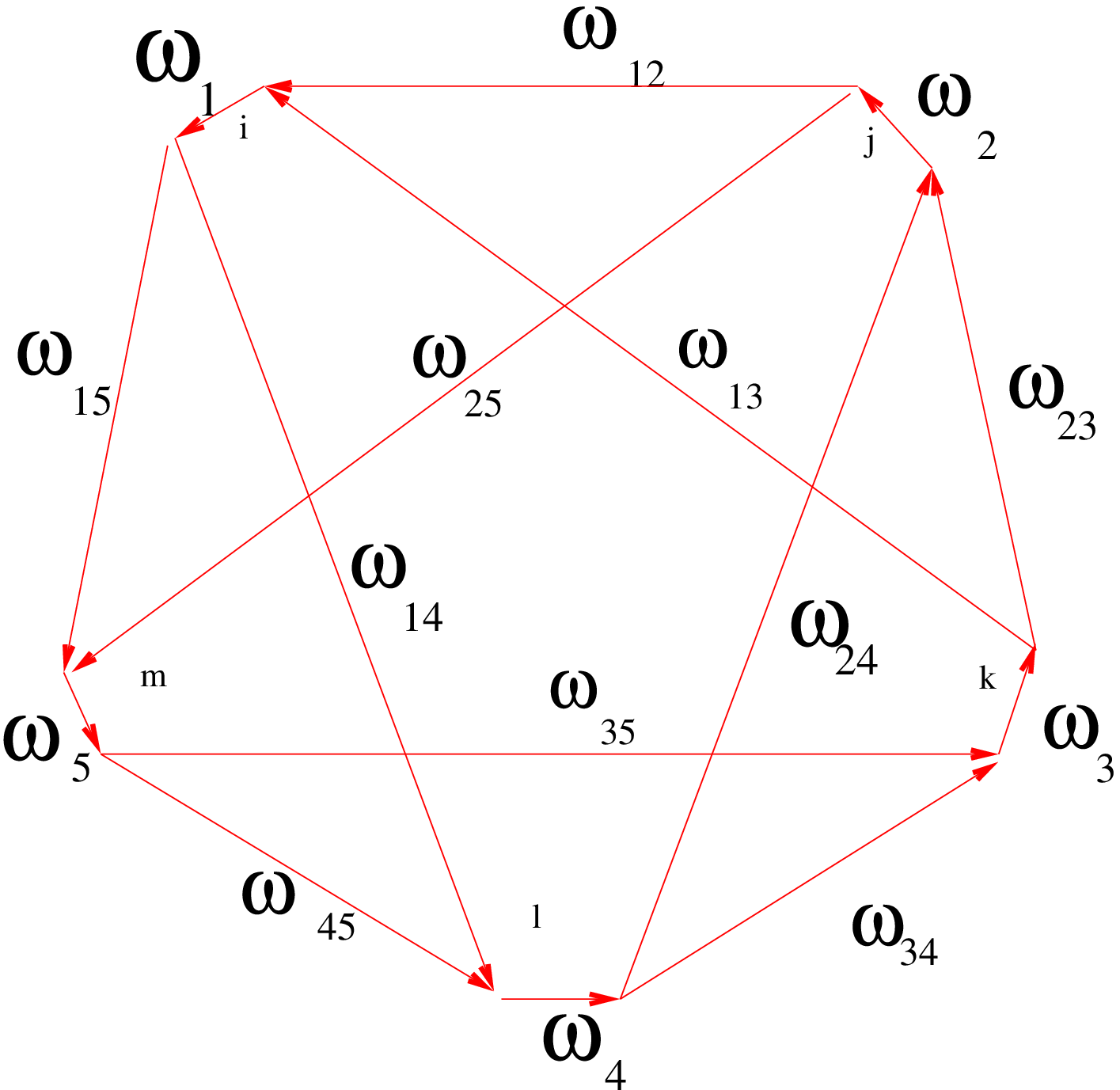}%
}%
.
\]
The final partition function is
\begin{equation}
Z_{BF}(\Delta)=\int\limits_{\{\omega_{b,}\omega_{e}\}}\prod_{b}\frac
{d_{\omega_{b}}}{64\pi^{8}}\prod_{s}Z_{BF}(s)\prod_{b}d\omega_{b}\prod
_{e}d\omega_{e}, \label{eq.6}%
\end{equation}
where the $Z_{BF}(s)=\{15\omega\}$ is the amplitude for a four-simplex $s$.
The $d_{\omega_{b}}=\left\vert \chi_{L}\chi_{R}\right\vert ^{2}$ term is the
quantum amplitude associated with the bone $b$.\ Here $\omega_{e}$ is the
internal representation used to define the intertwiners. Usually $\omega_{e}$
is replaced by $i_{e}$ to indicate the intertwiner. The set $\left\{
\boldsymbol{\omega}_{b,}\boldsymbol{\omega}_{e}\right\}  $ of all $\omega_{b}%
$'s and $\omega_{e}$'s is usually called a \textbf{coloring} of the bones and
the edges. This partition function may not be finite in general.

It is well known that the$\ BF$ theories are topological field theories. A
priori one cannot expect this to be true for the case of the BF spin foam
models because of the discretization of the BF action. For the spin foam
models of the BF\ theories for compact groups, it has been shown that the
partition functions are triangulation independent up to a factor
\cite{SpinFoamDiag}. This analysis is purely based on spin foam diagrammatics
and is independent of the group used as long the BF spin foam is defined
formally by equation (\ref{eq.del}) and the harmonic expansion in equation
(\ref{eq.del.exp}) is formally valid. So one can apply the spin foam
diagrammatics analysis directly to the $SO(4,C)$ BF spin foam and write down
the triangulation independent partition function as%

\[
Z_{BF}^{^{\prime}}(\Delta)=\tau^{n_{4}-n_{3}}Z_{BF}(\Delta)
\]
using the result from \cite{SpinFoamDiag}. In the above equation $n_{4},n_{3}$
is number of four bubbles and three bubbles in the triangulation $\Delta$ and
\begin{align*}
\tau &  =\delta_{SO(4,C)}(I)\\
&  =\frac{1}{64\pi^{8}}\int d_{\omega}^{2}d\omega.
\end{align*}
The above integral is divergent and so the partition functions need not be
finite. The normalized partition function is to be considered as the proper
partition function because the BF\ theory is supposed to be topological and so
triangulation independent.

\section{The $SO(4,C)$ Barrett-Crane Model}

\subsection{Classical $SO(4,C)$ General Relativity}

Consider a four dimensional manifold $M$. Let $A$ be a $SO(4,C)$ connection
1-form and $B^{ij}$ be a complex bivector valued $2$-form on $M$. I\ would
like to restrict myself to the non-degenerate general relativity in this
section by assuming $b=\frac{1}{4}\epsilon^{abcd}B_{ab}\wedge B_{cd}\neq0$.
The Plebanski action for the $SO(4,C)$ general relativity is obtained by
adding a Lagrange multiplier term to impose the Plebanski constraint to the BF
theory action given in equation :(\ref{BFaction}). A\ simple way of writing
the action \cite{MPR1} is%
\begin{equation}
S_{C}(A,B_{ij},\bar{A},\bar{B}_{ij},\phi)=\operatorname{Re}\left[  \int
_{M}tr(B\wedge F)+\frac{b}{2}\phi^{abcd}B_{ab}\wedge B_{cd}\right]  \text{,}
\label{GRactionComplex}%
\end{equation}
where $\phi$ is a complex tensor with the symmetries of the Riemann curvature
tensor such that $\phi^{abcd}\epsilon_{abcd}=0$. The field equations
corresponding to the extrema of the above action has been discussed by me in
\cite{ClassicalAreaReal}. Two important results are

\begin{itemize}
\item The Plebanski constraint imposes the condition $B_{ab}^{ij}=$
$\theta_{a}^{[i}\theta_{b}^{j]}$ where $\theta_{a}^{i}$ is a complex tetrad
field \cite{Plebanski}, \cite{ClassicalAreaReal}.

\item The field equations correspond to the $SO(4,C)$ general relativity on
the manifold $M$ \cite{ClassicalAreaReal}.
\end{itemize}

\subsubsection{Relation to Complex Geometry}

Let $M$ be a real analytic manifold. Let $M_{c}$ be the complex analytic
manifold which is obtained by analytically continuing the real coordinates on
$M$. The analytical continuation of the field equations and their solutions on
$M$ to complex $M_{c}$ can be used to define complex general relativity.
Conversely, the field equations of complex general relativity or their
solutions on $M_{c}$ when restricted to $M$ defines the $SO(4,C)$ general
relativity. Because of these properties the action $S$ can also be considered
as an action for complex general relativity.

Now consider the relation between the complex general relativity on $M_{c}$
and\ the $SO(4,C)$ general relativity on $M$. This relation critically depends
on $M$ being a real analytic manifold. It also depends on the fields on it
being analytic on some region may be except for some singularities. If the
fields and the field equations are discretized we lose the relation to complex
general relativity. Thus it is also not meaningful to relate a $SO(4,C)$
Barrett-Crane Model to complex general relativity. If the $SO(4,C)$
Barrett-Crane model has a semiclassical continuum general relativity limit
then a relation to complex general relativity may be recovered.

\subsection{The $SO(4,C)$ Barrett-Crane Constraints}

My goal here is to systematically construct the Barrett-Crane model of the
$SO(4,C)$ general relativity. In the previous section I\ discussed the
$SO(4,C)$ BF spin foam model. The basic elements of the BF spin foams are spin
networks built on graphs dual to the triangulations of the four simplices with
arbitrary intertwiners and the principal unitary representations of $SO(4,C)$
discussed in appendix B. These closed spin networks can be considered as
quantum states of four simplices in the BF\ theory and the essence of these
spin networks is mainly gauge invariance. To construct a spin foam model of
general relativity these spin networks need to be modified to include the
Plebanski Constraints in the discrete form.

A\ quantization of a four-simplex for the Riemannian general relativity was
proposed by Barrett and Crane \cite{BCReimmanion}. The bivectors $B_{i}$
associated with the ten triangles of a four-simplex in a flat Riemannian space
satisfy the following properties called the Barrett-Crane
constraints\footnote{I\ would like to refer the readers to the original paper
\cite{BCReimmanion} for more details.}:

\begin{enumerate}
\item The bivector changes sign if the orientation of the triangle is changed.

\item Each bivector is simple.

\item If two triangles share a common edge, then the sum of the bivectors is
also simple.

\item The sum of the bivectors corresponding to the edges of any tetrahedron
is zero. This sum is calculated taking into account the orientations of the
bivectors with respect to the tetrahedron.

\item The six bivectors of a four-simplex sharing the same vertex are linearly independent.

\item The volume of a tetrahedron calculated from the bivectors is real and non-zero.
\end{enumerate}

The items two and three can be summarized as follows:
\[
B_{i}\wedge B_{j}=0~\forall i,j,
\]
where $A\wedge B=\varepsilon_{IJKL}A^{IJ}B^{KL}$ and the $i,j$ represents the
triangles of a tetrahedron. If $i=j$, it is referred to as the simplicity
constraint. If $i\neq j$ it is referred as the cross-simplicity constraints.

Barrett and Crane have shown that these constraints are sufficient to restrict
a general set of ten bivectors $E_{b}$ so that they correspond to the
triangles of a geometric four-simplex up to translations and rotations in a
four dimensional flat Riemannian space.

The Barrett-Crane constraints theory can be trivially extended to the
$SO(4,C)$ general relativity. In this case the bivectors are complex and so
the volume calculated for the sixth constraint is complex. So we need to relax
the condition of the reality of the volume.

A quantum four-simplex for Riemannian general relativity is defined by
quantizing the Barrett-Crane constraints \cite{BCReimmanion}. The bivectors
$B_{i}$ are promoted to the Lie operators $\hat{B}_{i}$ on the representation
space of the relevant group and the Barrett-Crane constraints are imposed at
the quantum level. A\ four-simplex has been quantized and studied in the case
of the Riemannian general relativity before \cite{BCReimmanion}. All the first
four constraints have been rigorously implemented in this case. The last two
constraints are inequalities and they are difficult to impose. This could be
related to the fact that the Riemannian Barrett-Crane model reveal the
presence of degenerate sectors \cite{BaezEtalAsym}, \cite{JWBCS} in the
asymptotic limit \cite{JWBRW} of the model. For these reasons here after
I\ would like to refer to a spin foam model that satisfies only the first four
constraints as an \textit{essential Barrett-Crane model}, While a spin foam
model that satisfies all the six constraints as a \textit{rigorous
Barrett-Crane model}.

Here I\ would like to derive the essential $SO(4,C)$ Barrett-Crane model. For
this one must deal with complex bivectors instead of real bivectors. The
procedure that I\ would like to use to solve the constraints can be carried
over directly to the Riemannian Barrett-Crane\ model. This derivation
essentially makes the derivation of the Barrett-Crane intertwiners for the
real and the complex Riemannian general relativity more rigorous.

\subsubsection{The Simplicity Constraint}

The group $SO(4,C)$ is locally isomorphic to $\frac{SL(2,C)\times
SL(2,C)}{Z_{2}}$. An element $B$ of the Lie algebra space of $SO(4,C)$ can be
split into the left and the right handed $SL(2,C)$ components,%
\begin{equation}
B=B_{L}+B_{R}.
\end{equation}
There are two Casimir operators for $SO(4,C)$ which are $\varepsilon
_{IJKL}B^{IJ}B^{KL}$ and $\eta_{IK}\eta_{JL}B^{IJ}B^{KL}$, where $\eta_{IK}$
is the flat Euclidean metric. In terms of the left and right handed split I
can expand the Casimir operators as%
\[
\varepsilon_{IJKL}B^{IJ}B^{KL}=B_{L}\cdot B_{L}-B_{R}\cdot B_{R}~~\text{and}%
\]%
\[
\eta_{IK}\eta_{JL}B^{IJ}B^{KL}=B_{L}\cdot B_{L}+B_{R}\cdot B_{R},
\]
where the dot products are the trace in the $SL(2,C)$ Lie algebra coordinates.

The bivectors are to be quantized by promoting the Lie algebra vectors to Lie
operators on the unitary representation space of $SO(4,C)\approx$
$\frac{SL(2,C)\times SL(2,C)}{Z_{2}}$. The relevant unitary representations of
$SO(4,C)\simeq$ $SL(2,C)\otimes SL(2,C)/Z_{2}$ are labeled by a pair
($\chi_{L},$ $\chi_{R}$) such that $n_{L}+n_{R}$ is even (appendix B). The
elements of the representation space $D_{\chi_{L}}\otimes$ $D_{\chi_{R}}$ are
the eigen states of the Casimirs and on them the operators reduce to the
following:
\begin{equation}
\varepsilon_{IJKL}\hat{B}^{IJ}\hat{B}^{KL}=\frac{\chi_{L}^{2}-\chi_{R}^{2}}%
{2}\hat{I}~~\text{and} \label{eq.1}%
\end{equation}%
\begin{equation}
\eta_{IK}\eta_{JL}\hat{B}^{IJ}\hat{B}^{KL}=\frac{\chi_{L}^{2}+\chi_{R}^{2}%
-2}{2}\hat{I}. \label{eq.2}%
\end{equation}
The equation (\ref{eq.1}) implies that on $D_{\chi_{L}}\otimes$ $D_{\chi_{R}}$
the simplicity constraint $B\wedge B=0$ is equivalent to the condition
$\chi_{L}=\pm\chi_{R}$. I would like to find a representation space on which
the representations of $SO(4,C)$ are restricted precisely by $\chi_{L}$ $=$
$\pm\chi_{R}$. Since a $\chi$ representation is equivalent to $-\chi$
representations \cite{IMG}, $\chi_{L}=+\chi_{R}$ case is equivalent to
$\chi_{L}=-\chi_{R}$ \cite{IMG}.

Consider a square integrable function $f$ $(x)$ on the complex sphere $CS^{3}$
defined by%

\[
x\cdot x=1,\forall x\in\boldsymbol{C}^{4}.
\]
It can be Fourier expanded in the representation matrices of $SL(2,C)$ using
the isomorphism $CS^{3}\simeq$ $SL(2,C)$,%
\begin{equation}
f(x)=\frac{1}{8\pi^{4}}\int Tr(F(\chi)T_{\chi}(\mathfrak{g}(x)^{-1})\chi
\bar{\chi}d\chi, \label{Cs3Exp}%
\end{equation}
where the isomorphism $\mathfrak{g}\mathrm{:}CS^{3}$ $\longrightarrow SL(2,C)$
is defined in the appendix A. The group action of $g=(g_{L},g_{R})\in SO(4,C)$
on $x$ $\in CS^{3}$ is given by
\begin{equation}
\mathfrak{g}(gx)=g_{L}^{-1}\mathfrak{g}(x)g_{R}. \label{Cs3Action}%
\end{equation}
Using equation (\ref{Cs3Exp}) I can consider the $T_{\chi}(\mathfrak{g}%
(x))(z_{1},z_{2})$ as the basis functions of $L^{2}$ functions on $CS^{3}.$
The matrix elements of the action of $g$ on $CS^{3}$ is given by (appendix B)%
\[
\int\bar{T}_{\acute{\chi}}(\mathfrak{g}(x))(\acute{z}_{1},\acute{z}%
_{2})T_{\chi}(\mathfrak{g}(gx))(z_{1},z_{2})dx=T_{-\acute{\chi}}(g_{L}%
)(\acute{z}_{1},z_{1})T_{\chi}(g_{R})(\acute{z}_{2},z_{2})\delta(\acute{\chi
}-\chi).
\]
I see that the representation matrices are precisely those of $SO(4,C)$ only
restricted by the constraint $\chi_{L}=-\chi_{R}\approx\chi_{R}$. So the
simplicity constraint effectively reduces the Hilbert space $H$ to the space
of $L^{2}$ functions on $CS^{3}$. In Ref:\cite{BFfoamHighD} the analogous
result has been shown for $SO(N,R)$ where the Hilbert space is reduced to the
space of the $L^{2}$ functions on $S^{N-1}$.

\subsubsection{The Cross-simplicity Constraints}

Next let me quantize the cross-simplicity constraint part of the Barrett-Crane
constraint. Consider the quantum state space associated with a pair of
triangles $1$ and $2$ of a tetrahedron. A general quantum state that just
satisfies the simplicity constraints $B_{1}\wedge B_{1}=0$ and $B_{2}\wedge
B_{2}=0$ is of the form $f(x_{1},x_{2})$ $\in L^{2}(CS^{3}\ast CS^{3})$,
$x_{1},x_{2}\in CS^{3}$.

On the elements of $L^{2}(CS^{3}\ast CS^{3})$ the action $B_{1}\wedge B_{2}$
is equivalent to the action of $\left(  B_{1}+B_{2}\right)  \wedge\left(
B_{1}+B_{2}\right)  $\footnote{Please notice that
\[
\left(  \hat{B}_{1}+\hat{B}_{2}\right)  \wedge\left(  \hat{B}_{1}+\hat{B}%
_{2}\right)  =\hat{B}_{1}\wedge\hat{B}_{1}+\hat{B}_{2}\wedge\hat{B}_{2}%
+2B_{1}\wedge\hat{B}_{2}.
\]
But since $\hat{B}_{1}\wedge\hat{B}_{1}=\hat{B}_{2}\wedge\hat{B}_{2}=0$ on
$f(x_{1},x_{2})$ we have
\[
\left(  \hat{B}_{1}+\hat{B}_{2}\right)  \wedge\left(  \hat{B}_{1}+\hat{B}%
_{2}\right)  f(x_{1},x_{2})=\hat{B}_{1}\wedge\hat{B}_{2}f(x_{1},x_{2}).
\]
}. This implies that the cross-simplicity constraint $B_{1}\wedge B_{2}=0$
requires the simultaneous rotation of $x_{1},x_{2}$ involve only the $\chi
_{L}$ $=$ $\pm\chi_{R}$ representations. The simultaneous action of
$g=(g_{L},g_{R})$ on the arguments of $f(x_{1},x_{2})$ is%
\begin{equation}
gf(x_{1},x_{2})=f(g_{L}^{-1}x_{1}g_{R},g_{L}^{-1}x_{2}g_{R}). \label{action}%
\end{equation}
The harmonic expansion of $f(x_{1},x_{2})$ in terms of the basis function
$T_{\chi}(\mathfrak{g}(x))(z_{1},z_{2})$ is%
\[
f(x_{1},x_{2})=F_{z_{1}z_{2}\chi_{1}\chi_{2}}^{\acute{z}_{1}\acute{z}_{2}%
}T_{\acute{z}_{1}\chi_{1}}^{z_{1}}(\mathfrak{g}(x_{1}))T_{\acute{z}_{2}%
\chi_{2}}^{z_{2}}(\mathfrak{g}(x_{2})),
\]
where I have assumed all the repeated indices are either integrated or summed
over for equation only. The rest of the calculations can be understood
graphically. The last equation can be graphically written as follows:%

\[
f(x_{1},x_{2})=\iint\limits_{\chi_{1}\chi_{2}}%
\raisebox{-0.5613in}{\includegraphics[
height=0.9193in,
width=1.3318in
]%
{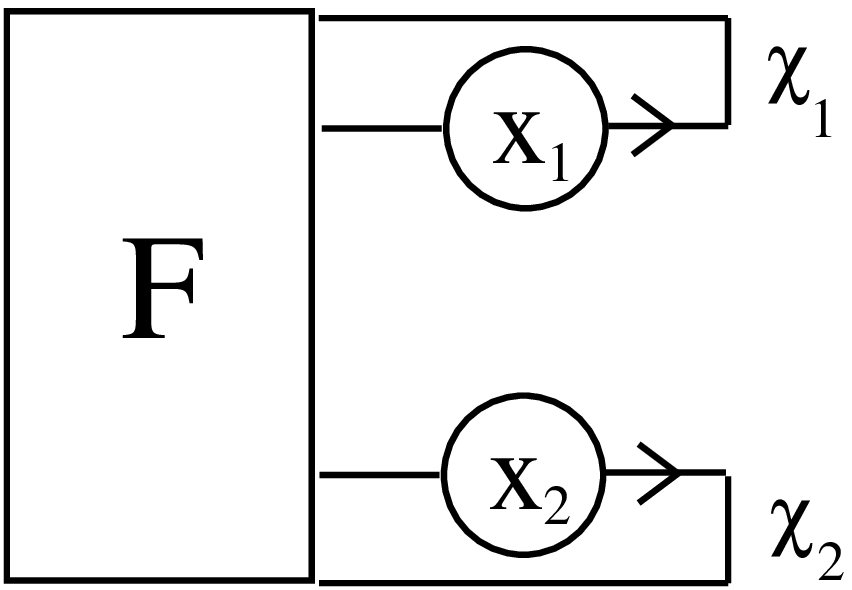}%
}%
d\chi_{1}d\chi_{2},
\]
where the box $F$ represents the tensor $F_{z_{1}z_{2}\chi_{1}\chi_{2}%
}^{\acute{z}_{1}\acute{z}_{2}}$. The action of $g\in SO(4,C)$ on $f$ is%

\begin{equation}
gf(x_{1},x_{2})=\iint\limits_{\chi_{1}\chi_{2}}%
\raisebox{-0.5518in}{\includegraphics[
trim=0.014528in 0.000000in -0.014528in 0.000000in,
height=1.0032in,
width=2.2935in
]%
{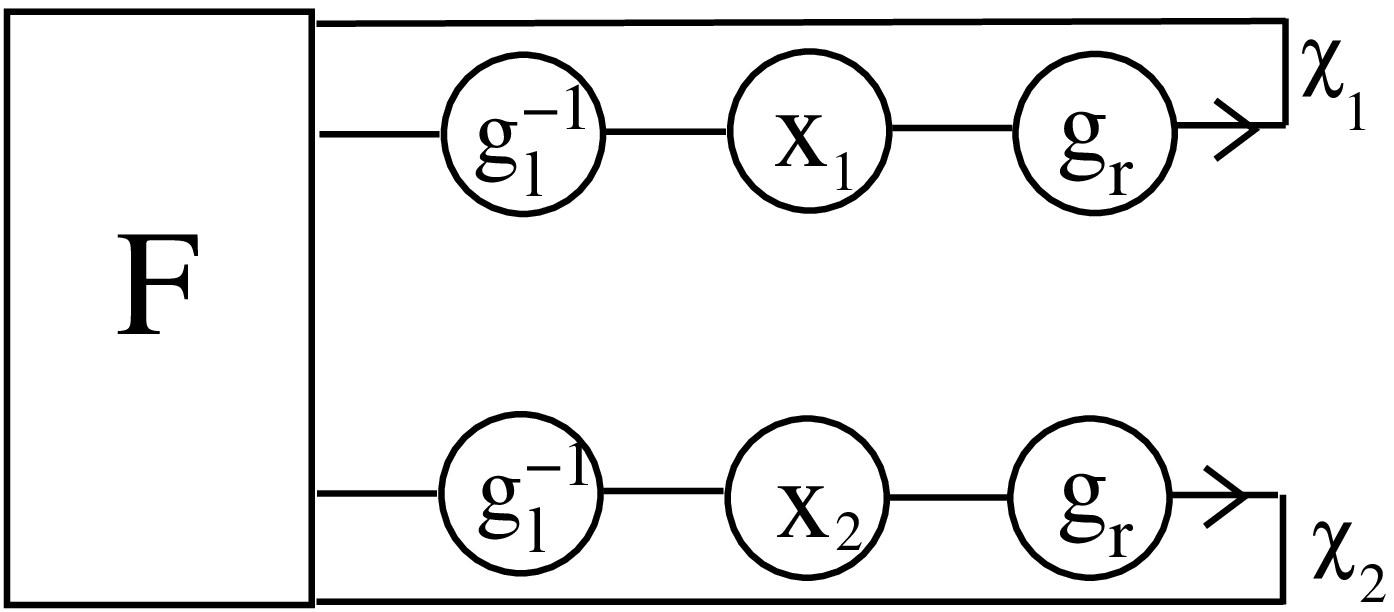}%
}%
d\chi_{1}d\chi_{2}. \label{gfresult}%
\end{equation}
Now for any $h\in SL(2,C),$%
\[
T_{a_{1}\chi_{1}}^{b_{1}}(h)T_{a_{2}\chi_{1}}^{b_{2}}(h)=C_{\chi_{1}\chi
_{2}b_{3}}^{b_{1}b_{2}\chi_{3}}\bar{C}_{a_{1}a_{2}\chi_{3}}^{\chi_{1}\chi
_{2}a_{3}}T_{a_{3}\chi_{3}}^{b_{3}}(h),
\]
where $C$'s are the Clebsch-Gordan coefficients of $SL(2,C)$ \cite{IMG},
\cite{NaimarckClebsch}\footnote{I\ derived this equation explicitly in the
appendix of Ref:\cite{MyRigorousSpinFoam}.}. I have assumed all the repeated
indices are either integrated or summed over for the previous and the next two
equations. Using this I can rewrite the $g_{L}$ and $g_{R}$ parts of the
result (\ref{gfresult}) as follows:%
\begin{equation}
T_{a_{1}\chi_{1}}^{z_{1}}(g_{L}^{-1})T_{a_{2}\chi_{2}}^{z_{2}}(g_{L}%
^{-1})=C_{\chi_{1}\chi_{2}z_{3}}^{z_{1}z_{2}\chi_{L}}\bar{C}_{a_{1}a_{2}%
\chi_{L}}^{\chi_{1}\chi_{2}a_{3}}T_{a_{3}\chi_{L}}^{z_{3}}(g_{L}^{-1})
\label{exp1}%
\end{equation}
and%
\begin{equation}
T_{\acute{z}_{1}\chi_{1}}^{b_{1}}(g_{R})T_{\acute{z}_{2}\chi_{2}}^{b_{2}%
}(g_{R})=C_{\chi_{1}\chi_{2}b_{3}}^{b_{1}b_{2}\chi_{R}}\bar{C}_{\acute{z}%
_{1}\acute{z}_{2}\chi_{R}}^{\chi_{1}\chi_{2}\acute{z}_{3}}T_{\acute{z}_{3}%
\chi_{R}}^{b_{3}}(g_{R}). \label{exp2}%
\end{equation}
Now we have
\[
gf(x_{1},x_{2})=\idotsint\limits_{\chi_{1}\chi_{2}\chi_{L}\chi_{R}}%
\raisebox{-0.5716in}{\includegraphics[
height=1.1588in,
width=2.2701in
]%
{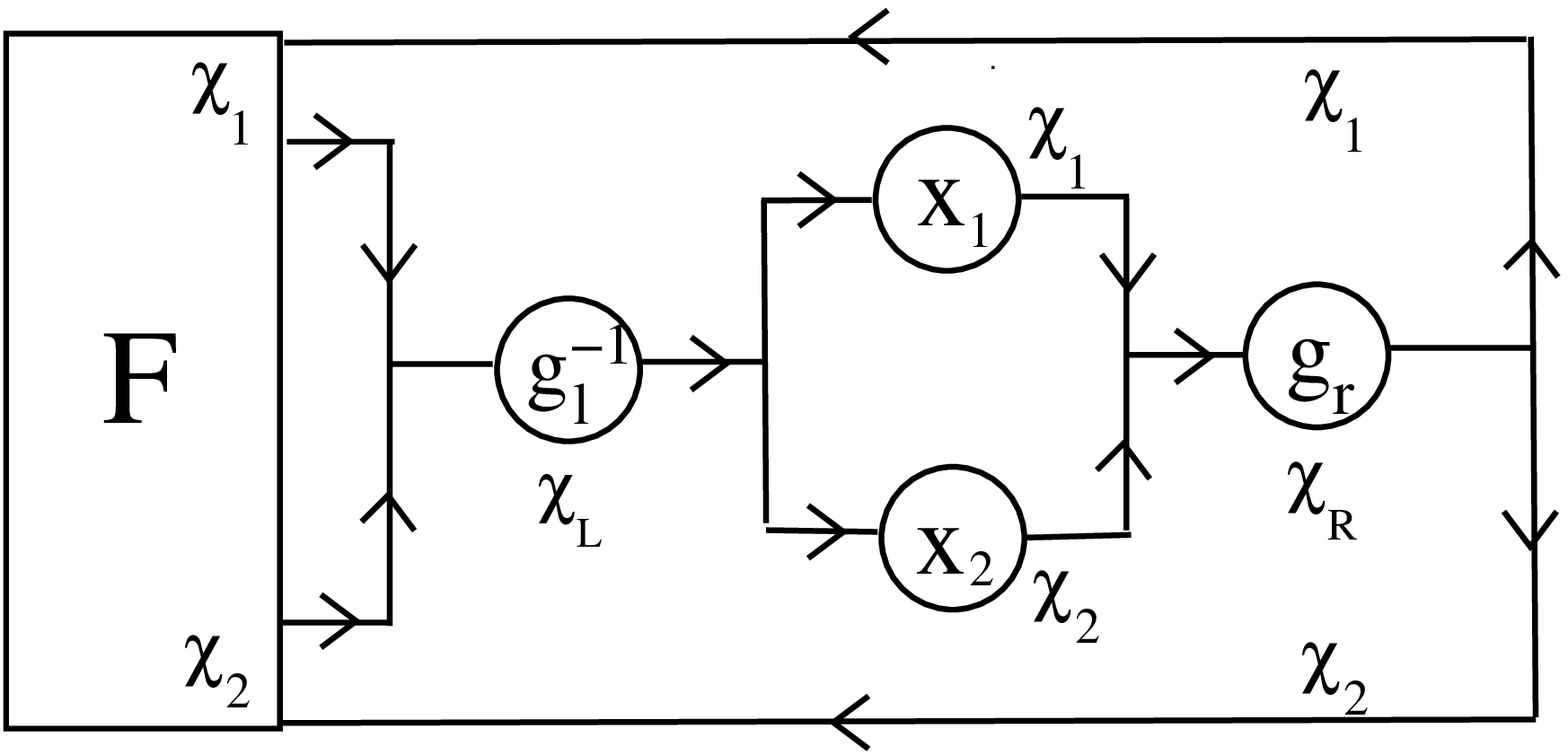}%
}%
d\chi_{1}d\chi_{2}.
\]

To satisfy the cross-simplicity constraint the expansion of $gf(x_{1},x_{2})$
must have contribution only from the terms with $\chi_{L}=\pm\chi_{R}$. In the
expansion in equation (\ref{exp1}) and equation (\ref{exp2}) in the right hand
side the terms are defined only up to a sign of $\chi_{L}$ and $\chi_{R}%
$\footnote{Please see appendix A for the explanation.}. Let me remove all the
terms which does not satisfy $\chi_{L}=\pm\chi_{R}$ (say = $\pm\chi$). Also
let me set $g=I.$ Now we can deduce that the functions denoted by $\tilde
{f}(x_{1},x_{2})$ obtained by reducing $f(x_{1},x_{2})$ using the
cross-simplicity constraints must have the expansion \footnote{The factor of 2
has been introduced to include the terms with $\chi_{L}=-\chi_{R}.$},%

\[
f(x_{1},x_{2})=2\iiint\limits_{\chi_{1}\chi_{2}\chi}c_{\chi}%
\raisebox{-0.5016in}{\includegraphics[
height=1.0006in,
width=2.2027in
]%
{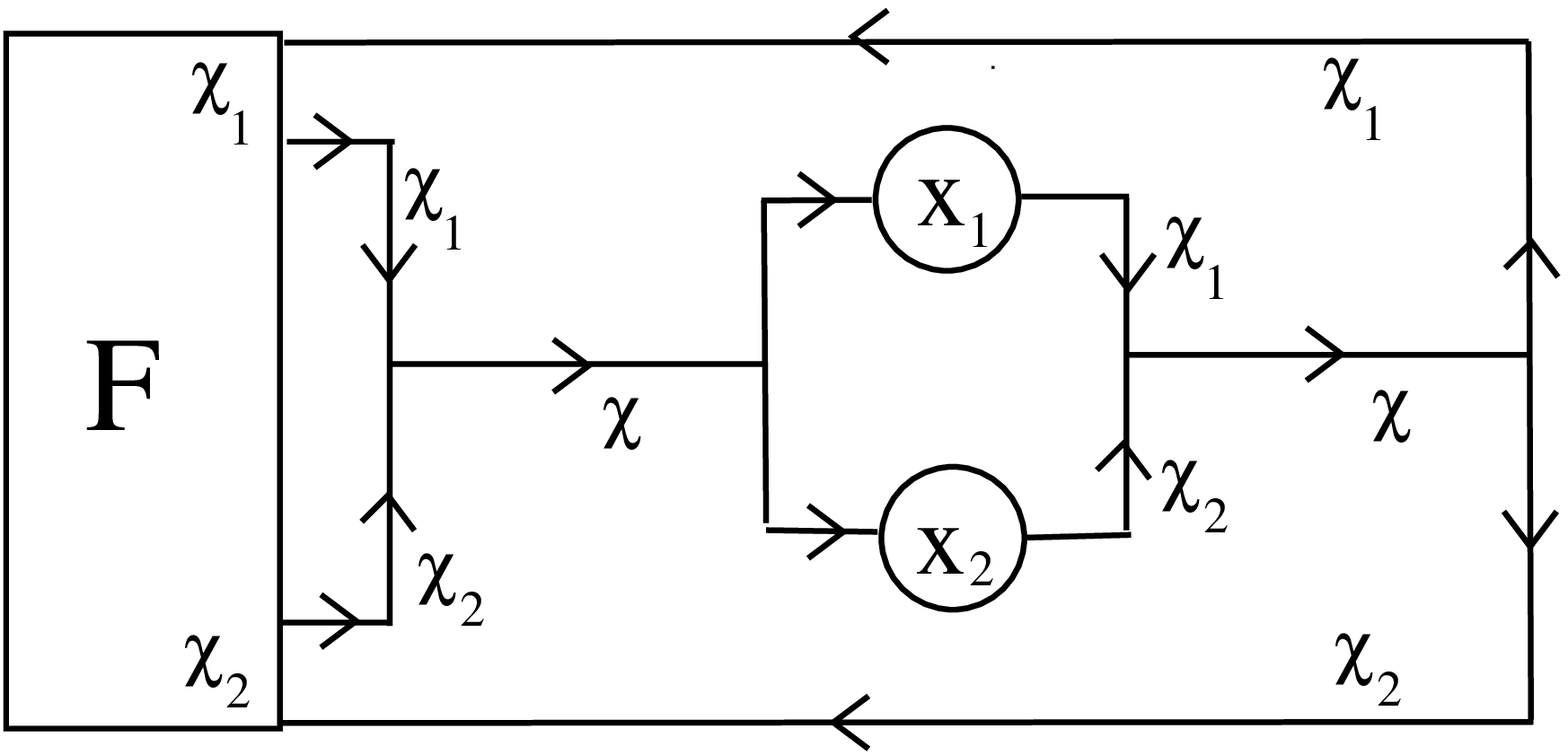}%
}%
d\chi_{1}d\chi_{2}d\chi,
\]
where the coefficients $c_{\chi}$ are arbitrary. Now the Clebsch-Gordan
coefficient terms in the expansion can be re-expressed using the following
equation :%
\begin{equation}
C_{\chi_{1}\chi_{2}z_{3}}^{z_{1}z_{2}\chi}\bar{C}_{\acute{z}_{1}\acute{z}%
_{2}\chi}^{\chi_{1}\chi_{2}\acute{z}_{3}}=\frac{8\pi^{4}}{\chi\bar{\chi}}%
\int_{SL(2,C)}T_{\acute{z}_{1}\chi_{1}}^{z_{1}}(h)T_{\acute{z}_{2}\chi_{2}%
}^{z_{2}}(h)\bar{T}_{z_{3}\chi}^{\acute{z}_{3}}(h)dh, \label{clebsmpl}%
\end{equation}
where $h$, $\tilde{h}$ $\in$ $SL(2,C)$ and $dh$ the bi-invariant measure on
$SL(2,C)$. Using this in the middle two Clebsch-Gordan coefficients of
$\tilde{f}(x_{1},x_{2})$ we get
\[
\tilde{f}(x_{1},x_{2})=2\iiint\limits_{\chi_{1}\chi_{2}\chi}\int
_{SL(2,C)}\frac{8\pi^{4}c_{\chi}}{\chi\bar{\chi}}%
\raisebox{-0.6028in}{\includegraphics[
height=1.0499in,
width=1.8983in
]%
{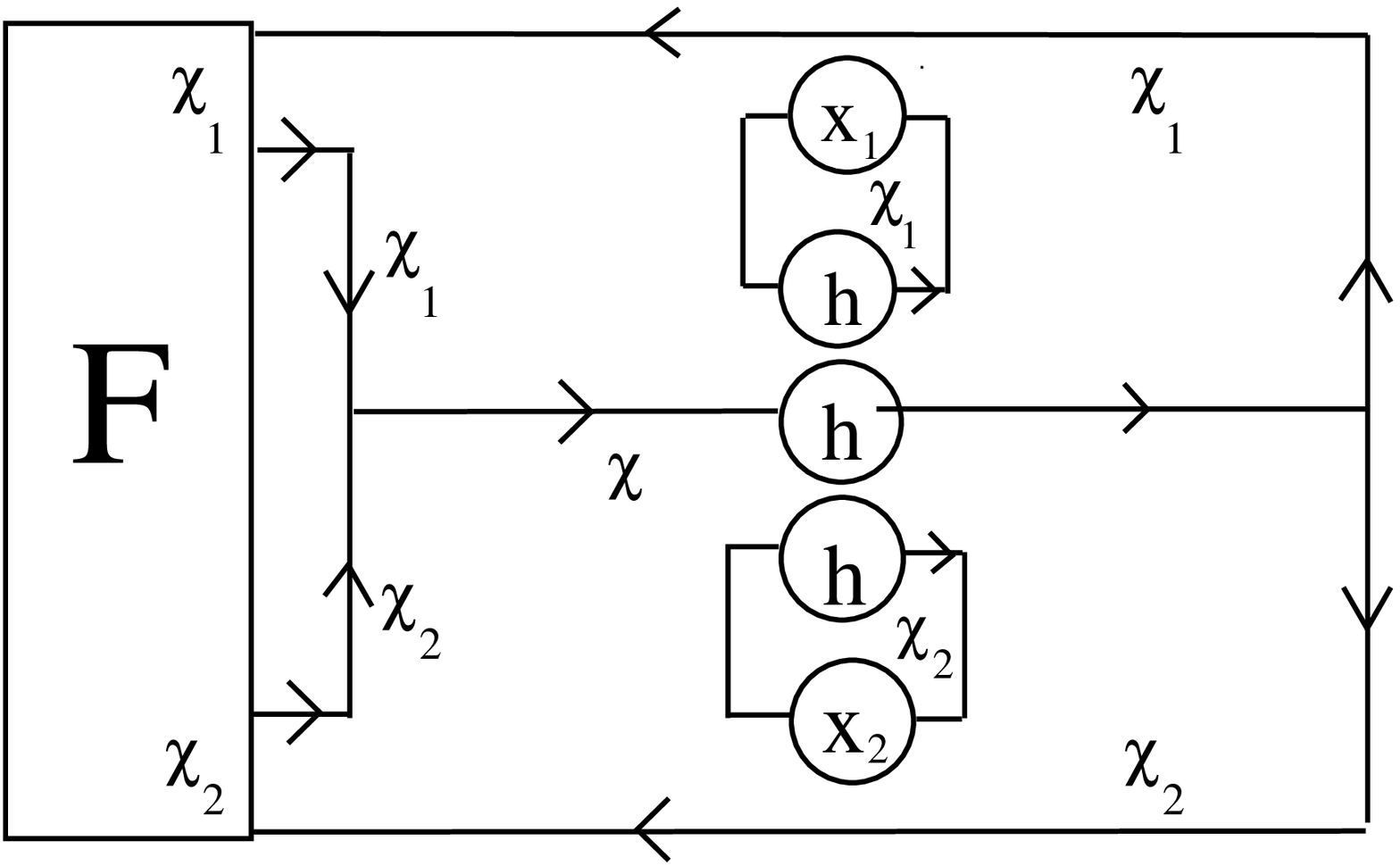}%
}%
dhd\chi_{1}d\chi_{2}d\chi.
\]
This result can be rewritten for clarity as%
\[
\tilde{f}(x_{1},x_{2})=2\iiint\limits_{\chi_{1}\chi_{2}\chi}\int
_{SL(2,C)}\frac{8\pi^{4}c_{\chi}}{\chi\bar{\chi}}%
\raisebox{-0.5016in}{\includegraphics[
height=1.0715in,
width=1.3214in
]%
{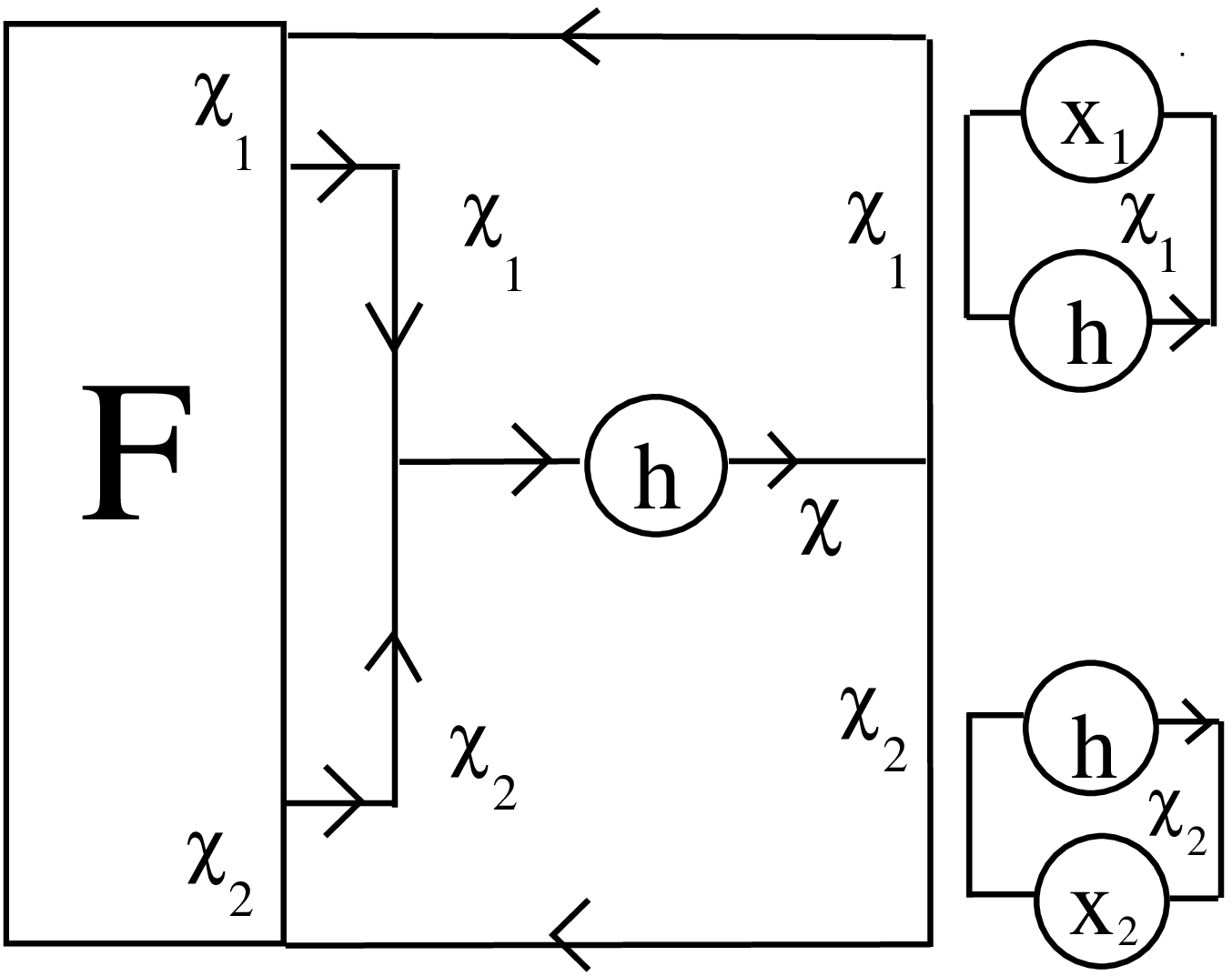}%
}%
dhd\chi_{1}d\chi_{2}d\chi.
\]
Once again applying equation (\ref{clebsmpl}) to the remaining two
Clebsch-Gordan coefficients we get,%

\[
\tilde{f}(x_{1},x_{2})=2\iiint\limits_{\chi_{1}\chi_{2}\chi}c_{\chi}%
\iint_{SL(2,C)\times SL(2,C)}\left(  \frac{8\pi^{4}}{\chi\bar{\chi}}\right)
^{2}%
\raisebox{-0.5016in}{\includegraphics[
height=1.0404in,
width=1.203in
]%
{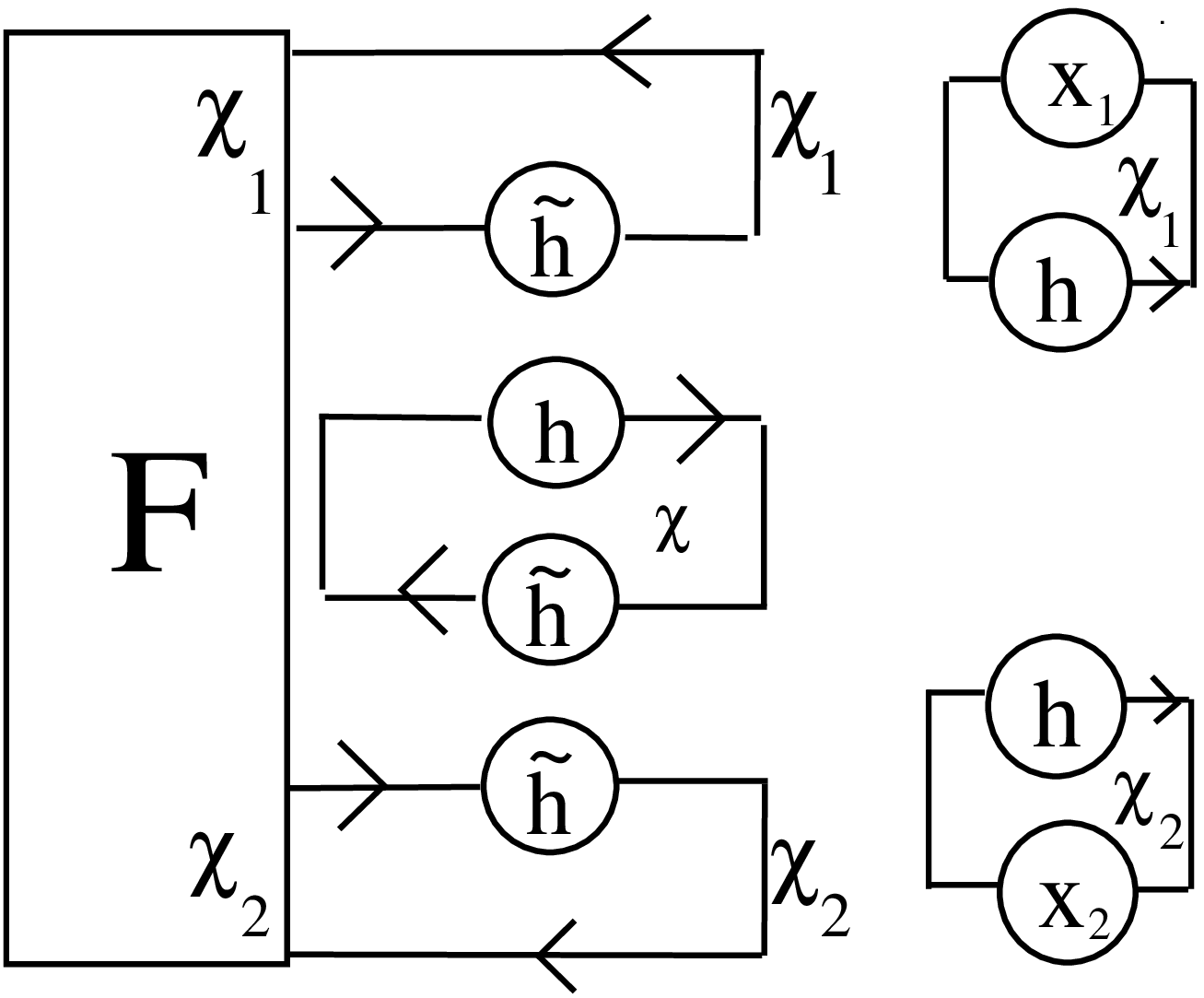}%
}%
dhd\tilde{h}d\chi_{1}d\chi_{2}d\chi.
\]

By rewriting the above expression, I deduce that a general function $\tilde
{f}(x_{1},x_{2})$ that satisfies the cross-simplicity constraint must be of
the form,
\begin{subequations}
\label{xsimpfunc}%
\begin{align*}
\tilde{f}(x_{1},x_{2})  &  =\iint\limits_{\chi_{1}\chi_{2}}c_{\chi}%
\int_{SL(2,C)}F_{\chi_{1}\chi_{2}}(h)%
\raisebox{-0.3684in}{\includegraphics[
height=0.7178in,
width=1.5368in
]%
{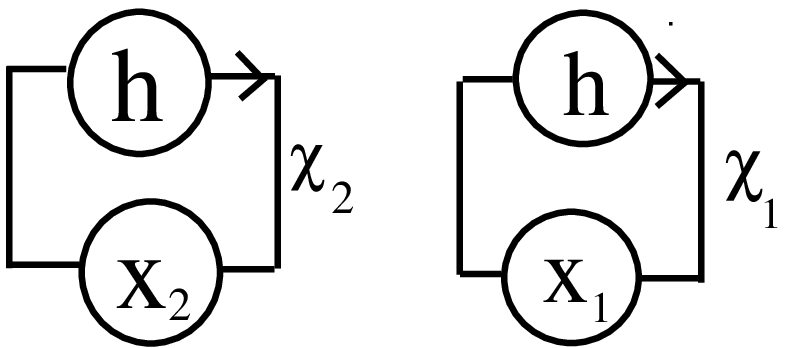}%
}%
dhd\chi_{1}d\chi_{2},\\
&  =\iint\limits_{\chi_{1}\chi_{2}}c_{\chi}\int_{SL(2,C)}F_{\chi_{1}\chi_{2}%
}(h)tr(T_{\chi_{1}}(\mathfrak{g}(x_{1})h)tr(T_{\chi_{2}}(\mathfrak{g}%
(x_{2})h)dhd\chi_{1}d\chi_{2},
\end{align*}
where $F_{\chi_{1}\chi_{2}}(h)$ is arbitrary. Then if $\Psi(x_{1},x_{2}%
,x_{3},x_{4})$ is the quantum state of a tetrahedron that satisfies all of the
simplicity constraints and the cross-simplicity constraints, it must be of the
form,
\end{subequations}
\begin{align*}
&  \Psi(x_{1},x_{2},x_{3},x_{4})\\
&  =\int F_{\chi_{1}\chi_{2}\chi_{3}\chi_{4}}(h)tr(T_{\chi_{1}}(\mathfrak{g}%
(x_{1})h)tr(T_{\chi_{2}}(\mathfrak{g}(x_{2})h)\\
&  tr(T_{\chi_{3}}(\mathfrak{g}(x_{3})h)tr(T_{\chi_{4}}(\mathfrak{g}%
(x_{4})h)dh\prod\limits_{i}d\chi_{i}.
\end{align*}
This general form is deduced by requiring that for every pair of variables
with the other two fixed, the function must be the form of the right hand side
of equation (\ref{xsimpfunc}).

\subsubsection{The $SO(4,C)$ Barrett-Crane Intertwiner}

Now the quantization of the fourth Barrett-Crane constraint demands that
$\Psi$ is invariant under the simultaneous complex rotation of its variables.
This is achieved if $F_{\chi_{1}\chi_{2}\chi_{3}\chi_{4}}(h)$ is constant
function of $h$. Therefore the quantum state of a tetrahedron is spanned by%
\begin{equation}
\Psi(x_{1},x_{2},x_{3},x_{4})=\int_{n\in CS^{3}}%
{\displaystyle\prod\limits_{i}}
T_{\chi_{i}}(\mathfrak{g}(x_{i})\mathfrak{g}(n))dn, \label{BCintertwiner}%
\end{equation}
where the measure $dn$ on $CS^{3}$ is derived from the bi-invariant measure on
$SL(2,C)$. I\ would like to refer to the functions $T_{\chi_{i}}%
(\mathfrak{g}(x_{i})$ as the $T-$\textbf{functions} here after.

\paragraph{Alternative forms}

The quantum state can be diagrammatically represented as follows:%
\[
\Psi(x_{1},x_{2},x_{3},x_{4})=\int%
\raisebox{-0.6599in}{\includegraphics[
height=1.3059in,
width=1.3059in
]%
{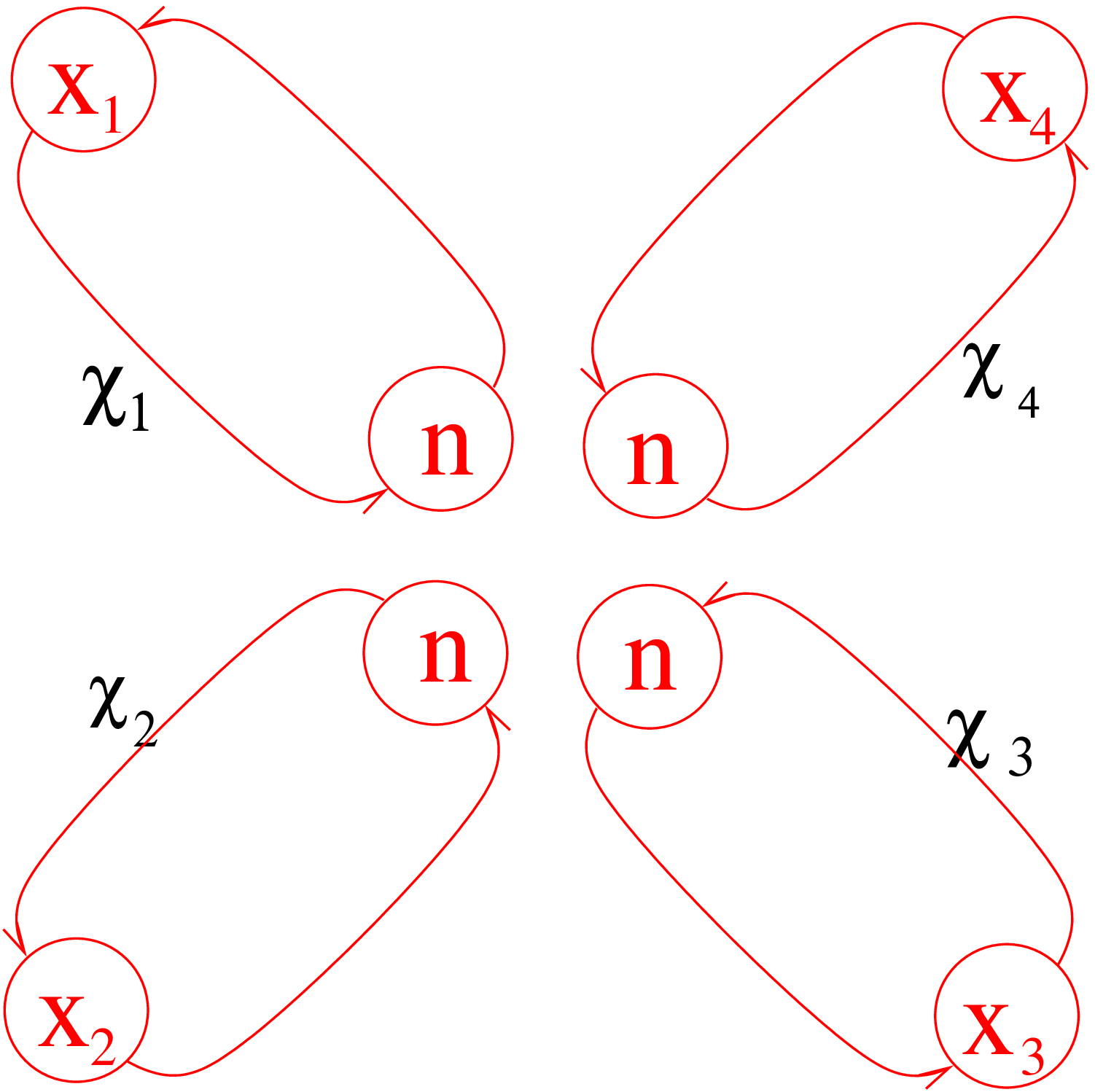}%
}%
dn.
\]
A unitary representation $T_{\chi}$ of $SL(2,C)$ can be considered as an
element of $D_{\chi}\otimes D_{\chi}^{\ast}$ where $D_{\chi}^{\ast}$ is the
dual representation of $D_{\chi}$. So using this the Barrett-Crane intertwiner
can be written as an element $\left\vert \Psi\right\rangle \in\bigotimes
\limits_{i}D_{\chi_{i}}\otimes D_{\chi_{i}}^{\ast}$ as follows:%
\[
\left\vert \Psi\right\rangle =\int\limits_{CS^{3}}%
\raisebox{-0.6529in}{\includegraphics[
height=1.2955in,
width=1.3474in
]%
{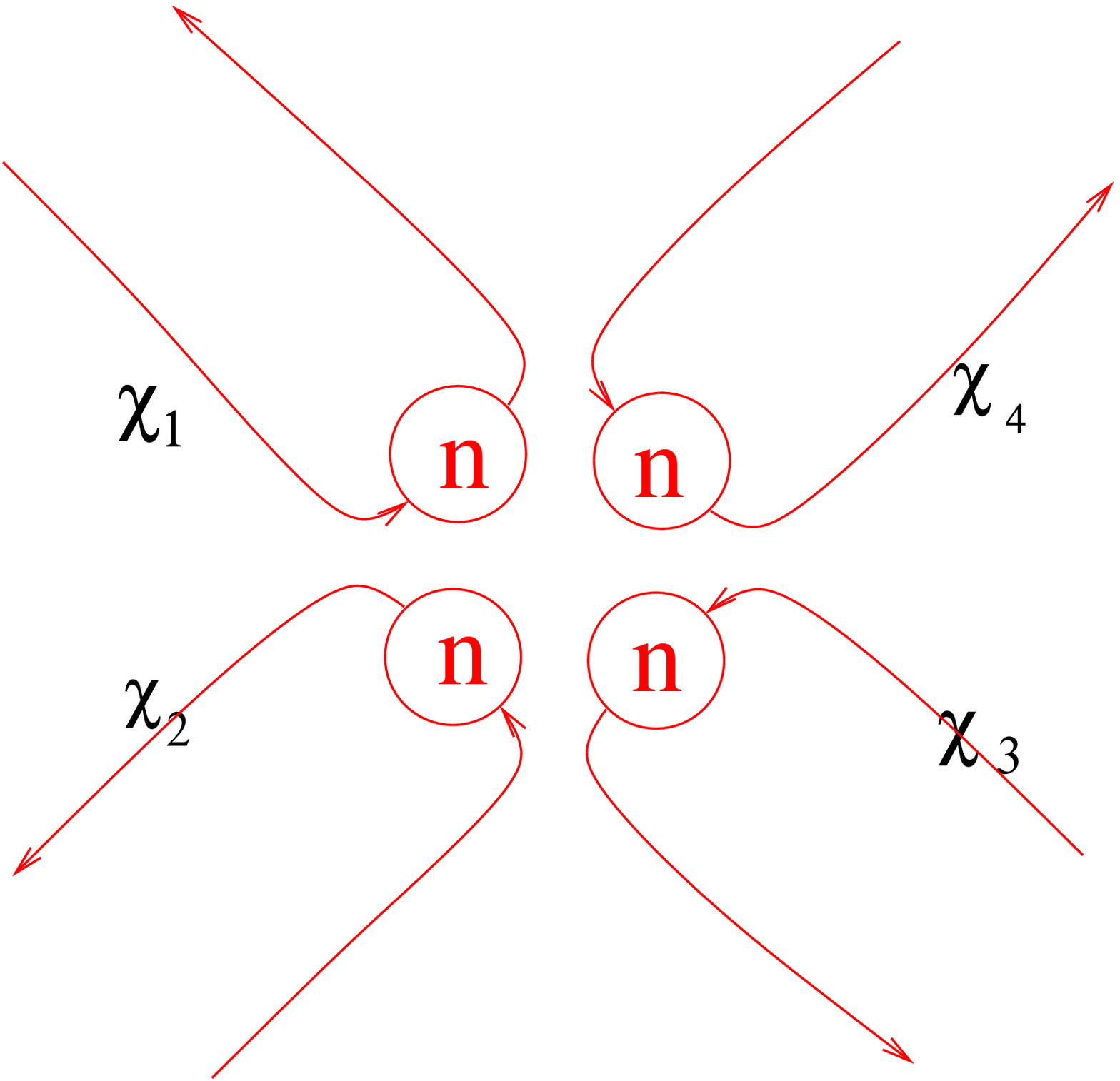}%
}%
dn.
\]
Since $SL(2,C)\approx CS^{3},$ using the following graphical identity:%

\[
\int_{SL(2,C)}%
\raisebox{-0.5716in}{\includegraphics[
height=1.1338in,
width=0.9824in
]%
{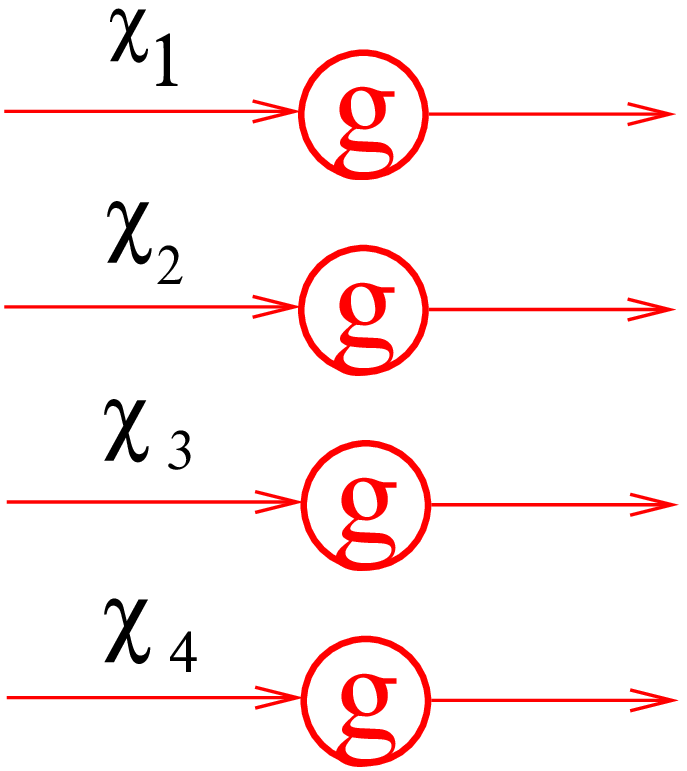}%
}%
dg=\int%
\raisebox{-0.5613in}{\includegraphics[
height=1.1122in,
width=2.0877in
]%
{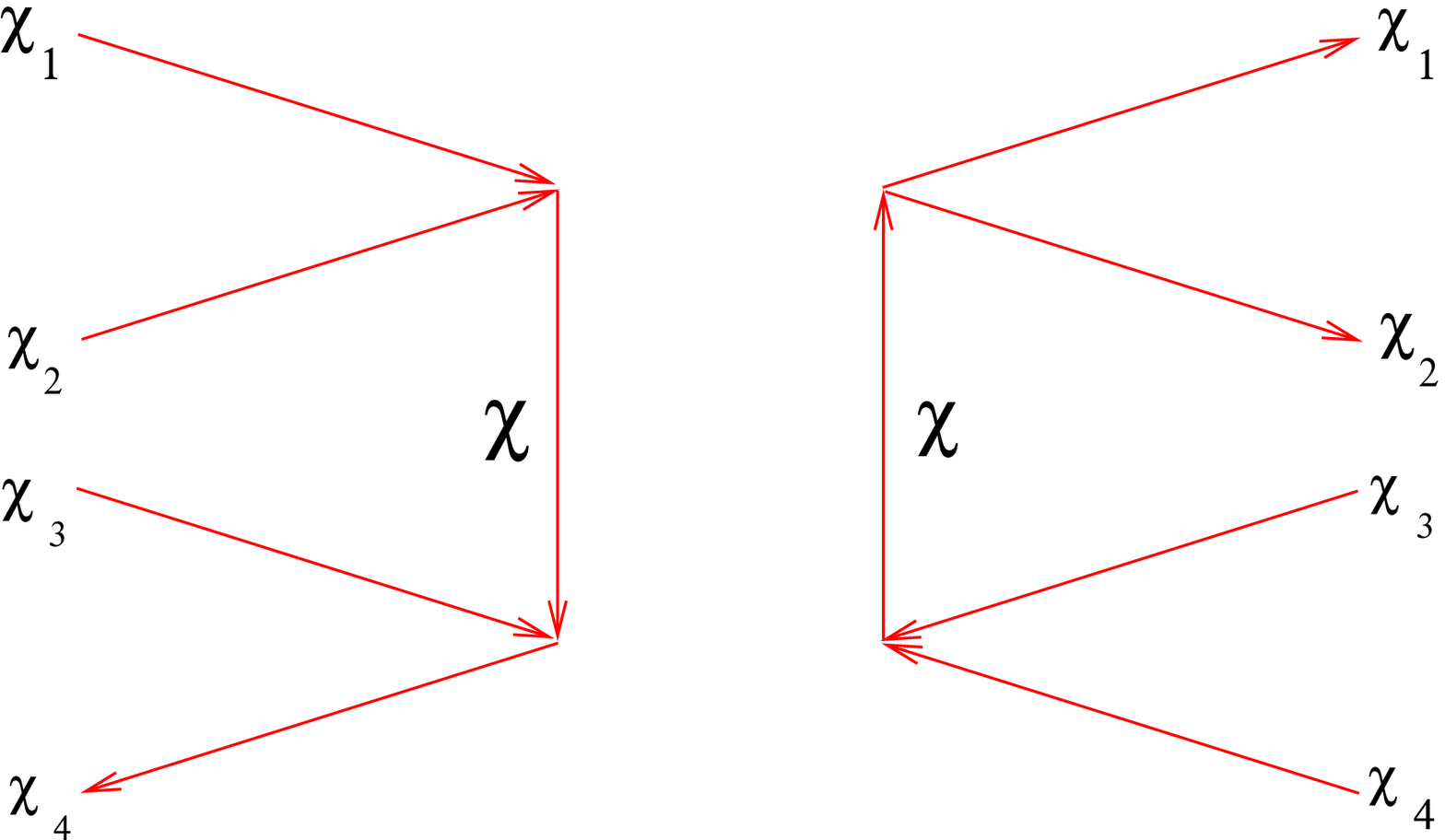}%
}%
\frac{8\pi^{4}}{\chi\bar{\chi}}d\chi,
\]
the Barrett-Crane solution can be rewritten as%

\[
\left\vert \Psi\right\rangle =\int%
\raisebox{-0.4817in}{\includegraphics[
height=0.9617in,
width=2.194in
]%
{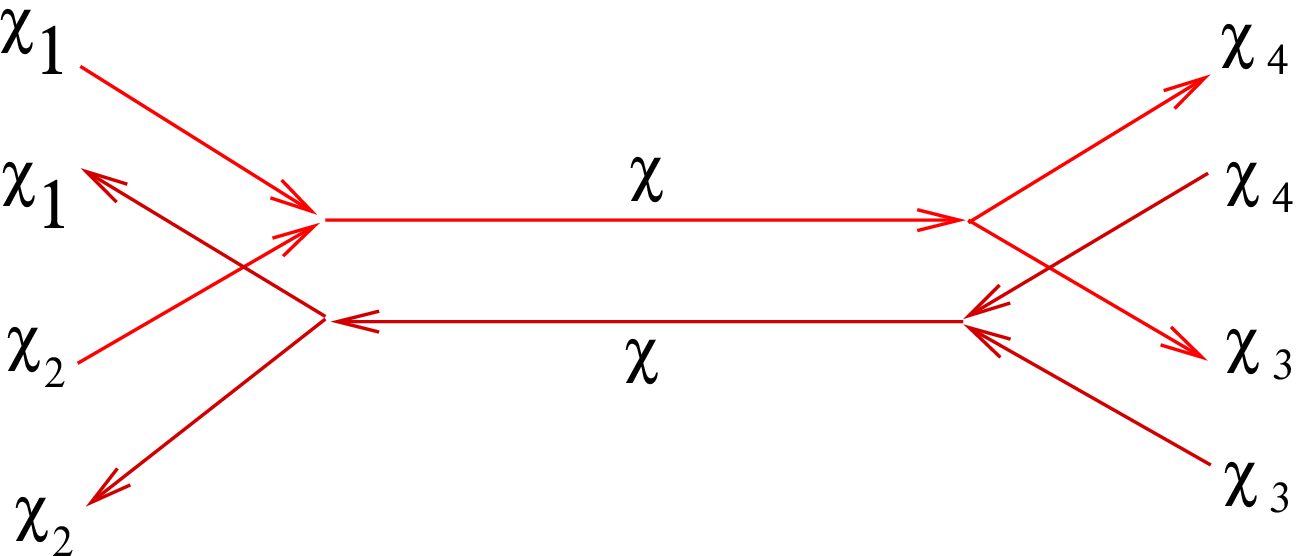}%
}%
\frac{8\pi^{4}}{\chi\bar{\chi}}d\chi,
\]
which emerges as an intertwiner in the familiar form in which Barrett and
Crane proposed it for the Riemannian general relativity. It can be clearly
seen that the simple representations for $SO(4,R)$ ($J_{L}=$ $J_{R}$) has been
replaced by the simple representation of $SO(4,C)$ ($\chi_{L}=$ $\pm\chi_{R}$).

\paragraph{Relation to the Riemannian Barrett-Crane model:}

All the analysis done until for the $SO(4,C)\ $Barrett-Crane theory can be
directly applied to the Riemannian$\ $Barrett-Crane theory. The
correspondences between the two models are listed in the following
table\footnote{BC stands for Barrett-Crane. For $\chi_{L}$ and $\chi_{R}$ we
have $n_{L}+n_{R}=even$.
\par
{}}:

\begin{center}%
\begin{tabular}
[c]{lll}%
\textbf{Property} & $SO(4,R)$ BC model & $SO(4,C)$ BC model\\
Gauge group & $SO(4,R)\approx\frac{SL(2,C)\otimes SL(2,C)}{Z_{2}}$ &
$SO(4,C)\approx\frac{SU(2)\otimes SU(2)}{Z_{2}}$\\
Representations & $J_{L},J_{R}$ & $\chi_{L},\chi_{R}$\\
Simple representations & $J_{L}=J_{R}$ & $\chi_{L}=\pm\chi_{R}$\\
Homogenous space & $S^{3}\approx SU(2)$ & $CS^{3}\approx SL(2,C)$%
\end{tabular}

\end{center}

\subsubsection{The Spin Foam Model for the $SO(4,C)$ General Relativity.}

The $SO(4,C)$ Barrett-Crane intertwiner derived in the previous section can be
used to define a $SO(4,C)$ Barrett-Crane spin foam model. The amplitude
$Z_{BC}(s)$ of a four-simplex $s$ is given by the $\{10\chi\}_{SO(4,C)}$
symbol given below:%

\begin{equation}
\{10\chi\}_{SO(4,C)}=%
\raisebox{-0.7022in}{\includegraphics[
height=1.4131in,
width=1.5126in
]%
{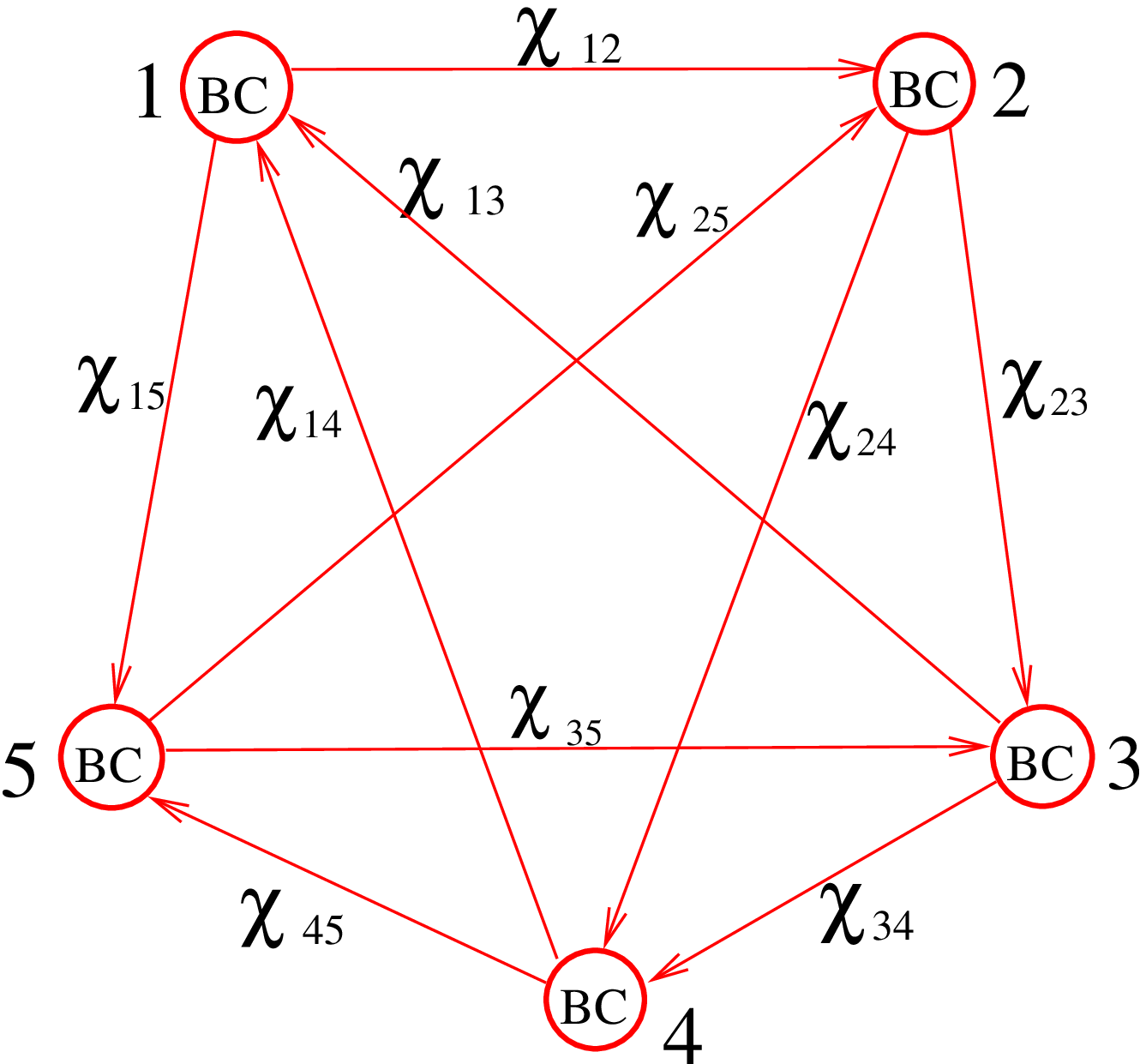}%
}%
\text{,} \label{10w}%
\end{equation}
where the circles are the Barrett-Crane intertwiners. The integers represent
the tetrahedra and the pairs of integers represent triangles. The intertwiners
use the four $\chi$'s associated with the links that emerge from it for its
definition in equation (\ref{10w}). In the next subsection, the propagators of
this theory are defined and the $\{10\chi\}$ symbol is expressed in terms of
the propagators in the subsubsection that follows it.

The $SO(4,C)$ Barrett-Crane partition function of the spin foam associated
with the four dimensional simplicial manifold with a triangulation $\Delta$
is
\begin{equation}
Z(\Delta)=\sum_{\left\{  \chi_{b}\right\}  }\left(  \prod_{b}\frac{d_{\chi
_{b}}^{2}}{64\pi^{8}}\right)  \prod_{s}Z(s), \label{PartitionFunction}%
\end{equation}
where $Z(s)$ is the quantum amplitude associated with the $4$-simplex $s$ and
the $d_{\chi_{b}}$ adopted from the spin foam model of the $BF\ $theory can be
interpreted as the quantum amplitude associated with the bone $b$.

\subsubsection{The Features of the $SO(4,C)$ Spin Foam}

\begin{itemize}
\item Areas: The squares of the areas of the triangles (bones) of the
triangulation are given by $\eta_{IK}\eta_{JL}B^{IJ}B^{KL}$. The eigen values
of the squares of the areas in the $SO(4,C)$ Barrett-Crane model from equation
(\ref{eq.2}) are given by
\begin{align*}
\eta_{IK}\eta_{JL}\hat{B}_{b}^{IJ}\hat{B}_{b}^{KL}  &  =\left(  \chi
^{2}-1\right)  \hat{I}\\
&  =\left(  \frac{n^{2}}{2}-\rho^{2}-1+i\rho n\right)  \hat{I}.
\end{align*}
One can clearly see that the area eigen values are complex. The $SO(4,C)$
Barrett-Crane model relates to the $SO(4,C)$ general relativity. Since in the
$SO(4,C)$ general relativity the bivectors associated with any two dimensional
flat object are complex, it is natural to expect that the areas defined in
such a theory are complex too. This is a generalization of the concept of the
space-like and the time-like areas for the real general relativity models:
Area is imaginary if it is time-like and real if it is space-like.

\item Propagators: Laurent and Freidel have investigated the idea of
expressing simple spin networks as Feynman diagrams \cite{SimpleSpinNetworks}.
Here we will apply this idea to the $SO(4,C)$ simple spin networks. Let
$\Sigma$ be a triangulated three surface. Let $n_{i}\in CS^{3}$ be a vector
associated with the $i^{th}$ tetrahedron of the $\Sigma$. The propagator of
the $SO(4,C)$ Barrett-Crane model associated with the triangle $ij$ is given
by%
\begin{align*}
G_{\chi_{ij}}(n_{i},n_{j})  &  =Tr(T_{\chi_{ij}}(g(n_{i}))T_{\chi_{ij}}^{\dag
}(g(n_{j})))\\
&  =Tr(T_{\chi_{ij}}(g(n_{i})g^{-1}(n_{j}))),
\end{align*}
where $\chi_{ij}$ is a representation associated with the triangle common to
the $i^{th}$ and the $j^{th}$ tetrahedron of $\Sigma$. If $X$ and $Y$ belong
to $CS^{3}$ then
\[
tr\left(  \mathfrak{g}(X)\mathfrak{g}(Y)^{-1}\right)  =2X.Y,
\]
where $X.Y$ is the Euclidean dot product and $tr$ is the matrix trace. If
$\lambda=e^{t}$ and $\frac{1}{\lambda}$ are the eigen values of $g(X)g(Y)^{-1}%
$ then,%
\begin{align*}
\lambda+\lambda^{-1}  &  =2X.Y\\
X.Y  &  =\cosh(t).
\end{align*}
From the expression for the trace of the $SL(2,C)$ unitary representations,
(appendix A, \cite{IMG}) I have the propagator for the $SO(4,C)$ Barrett-Crane
model calculated as%
\[
G_{\chi_{ij}}(n_{i},n_{j})=\dfrac{\cos(\rho_{ij}\eta_{ij}+n_{ij}\theta_{ij}%
)}{\left\vert \sinh(\eta_{ij}+i\theta_{ij})\right\vert ^{2}},
\]
where $\eta_{ij}+i\theta_{ij}$ is defined by $n_{i}.n_{j}=\cosh(\eta
_{ij}+i\theta_{ij})$. Two important properties of the propagators are listed below.

\begin{enumerate}
\item Using the expansion for the delta on $SL(2,C)$ I have
\begin{align*}
\delta_{CS^{3}}(X,Y)  &  =\delta_{SL(2,C)}(g(X)g^{-1}(Y))\\
&  =\frac{1}{8\pi^{4}}\int\bar{\chi}\chi Tr(T_{\chi}(g(X)g^{-1}(Y))d\chi,
\end{align*}
where the suffix on the deltas indicate the space in which it is defined.
Therefore%
\[
\int\bar{\chi}\chi G_{\chi}(X,Y))=8\pi^{4}\delta_{CS^{3}}(X,Y).
\]

\item Consider the orthonormality property of the principal unitary
representations of $SL(2,C)$ given by%
\begin{align*}
&  \int_{CS^{3}}T_{\acute{z}_{1}\chi_{1}}^{z_{1}}(g(X))T_{\acute{z}_{2}%
\chi_{2}}^{\dag z_{2}}(g(X))dX\\
&  =\frac{8\pi^{4}}{\chi_{1}\bar{\chi}_{1}}\delta(\chi_{1}-\chi_{2}%
)\delta(z_{1}-\acute{z}_{1})\delta(z_{2}-\acute{z}_{2}),
\end{align*}
where the delta on the $\chi$'s is defined up to a sign of them$.$ From this I
have%
\[
\int_{CS^{3}}G_{\chi_{1}}(X,Y)G_{\chi_{2}}(Y,Z)dY=\frac{8\pi^{4}}{\chi_{1}%
\bar{\chi}_{1}}\delta(\chi_{1}-\chi_{2})G_{\chi_{1}}(X,Z).
\]

\end{enumerate}

\item The $\{10\chi\}$ symbol can be defined using the propagators on the
complex three sphere as follows:%
\begin{align*}
Z(s)  &  =\int_{x_{k}\in CS^{3}}%
{\displaystyle\prod\limits_{i<j}}
T_{\chi_{ij}}(\mathfrak{g}(x_{i})\mathfrak{g}(x_{j}))%
{\displaystyle\prod\limits_{k}}
dx_{k},\\
&  =\int_{\forall x_{k}\in CS^{3}}%
{\displaystyle\prod\limits_{i<j}}
G_{\chi_{ij}}(x_{i}\mathfrak{,}x_{j})%
{\displaystyle\prod\limits_{k}}
dx_{k},
\end{align*}
where $i$ denotes a tetrahedron of the four-simplex. For each tetrahedron
$k,~$a free variable $x_{k}\in CS^{3}$ is associated. For each triangle $ij$
which is the intersection of the $i$'th and the $j$'th tetrahedron, a
representation of $SL(2,C)$ denoted by $\chi_{ij}$ is associated.

\item Discretization Dependence and Local Excitations: It is well known that
the BF theory is discretization independent and is topological. The$\ $spin
foam for the $SO(4,C)$ general relativity is got by imposing the Barrett-Crane
constraints on the $BF\ $Spin foam. After the imposition of the Barrett-Crane
constraints the theory loses the discretization independence and the
topological nature. This can be seen in many ways.

\begin{itemize}
\item The simplest reason is that the $SO(4,C)\ $Barrett-Crane model
corresponds to the quantization of the discrete $SO(4,C)$ general relativity
which has local degrees of freedom.

\item After the restriction of the representations involved in BF spin foams
to the simple representations and the intertwiners to the Barrett-Crane
intertwiners, various important identities used in the spin foam diagrammatics
and proof of the discretization independence of the BF\ theory spin foams in
Ref:\cite{SpinFoamDiag} are no longer available.

\item The BF partition function is simply gauge invariant measure of the
volume of space of flat connections. Consider the following harmonic expansion
of the delta function which was used in the derivation of the $SO(4,C)$ BF
theory:%
\[
\delta(g)=\frac{1}{8\pi^{4}}\int d_{\omega}tr(T_{\omega}(g))d\omega.
\]
Imposition of the Barrett-Crane constraints on the BF theory spin foam,
suppresses the terms corresponding to the non-simple representations. If only
the simple representations are allowed in the right hand side, it is no longer
peaked at the identity. This means that the partition function for
the\ $SO(4,C)$\ Barrett-Crane model involves contributions only from the
non-flat connections which has local information.

\item In the asymptotic limit study of the $SO(4,C)$ spin foams in section
four the discrete version of the $SO(4,C)$ general relativity (Regge calculus)
is obtained. The Regge calculus action is clearly discretization dependent and non-topological.
\end{itemize}

\item The real Barrett-Crane models that are discussed in the next section are
the restricted form of the $SO(4,C)\ $Barrett-Crane model. The above reasoning
can be applied to argue that they are also discretization dependent.
\end{itemize}

\section{Spin Foams for Real General Relativity}

\subsection{The Formal Structure of Barrett-Crane Intertwiners}

Let me briefly discuss the formal structure of the Barrett-Crane intertwiner
of the $SO(4,C)$ general relativity for the purpose of the developing spin
foam models for real general relativity theories. It has the following elements:

\begin{itemize}
\item A gauge group $G,$

\item A homogenous space $X$ of $G,$

\item A $G$ invariant measure on $X$ and,

\item A complete orthonormal set of functions which call as $T-$functions
which are maps from $X$ to the Hilbert spaces of a subset of unitary
representations of $G$:
\[
T_{\rho}:X\rightarrow D_{\rho},
\]
where $\rho$ is a representation of $G.$ The $T-$functions correspond to the
various unitary representations under the transformation of $X$ under $G.$ The
$T-$functions are complete in the sense that on the $L^{2}$ functions on $X$
they define invertible Fourier transforms.
\end{itemize}

Formally Barrett-Crane intertwiners are quantum states $\Psi$ associated to
closed simplicial two surfaces defined as an integral of a outer product of
$T-$functions on the space $X$:
\[
\Psi=\int\limits_{X}\prod\limits_{\otimes\rho}T_{\rho}(x)d_{X}x\in
\prod\limits_{\otimes\rho}D_{\rho}.
\]
It can seen that $\Psi$ is gauge invariant under $G$ because of the invariance
of the measure $d_{X}x$.

\subsection{The Real Barrett-Crane Models}

In the case of non-degenerate general relativity the reality of the area
metric is the necessary and sufficient condition for real geometry
\cite{ClassicalAreaReal}. In the Plebanski formalism the area metric can be
expressed in terms of the 2-form bivector field variable $B_{ab}^{IJ}$
$dx^{a}\wedge dx^{b}$ as $\eta_{IK}\eta_{JL}B^{IJ}\wedge B^{KL}$ where the
$\wedge$ represents the exterior product on the forms. It has been shown in
case of Ref:(\cite{ClassicalAreaReal}) that by adding a Lagrange multiplier to
the $SO(4,C)$ Plebanski action, we can derive real general relativity.

On a simplicial manifold a bivector two form field can be discretized by
associating bivectors to the triangles. The discrete equivalent of the area
metric reality constraint is the bivector scalar product reality constraint
\cite{ClassicalAreaReal}. Consider a four-simplex with complex bivectors
$B_{i}$, $i=1$ to $10$ associated with its triangles. Then the bivector scalar
product reality constraint requires%
\[
\operatorname{Im}(B_{i}\wedge B_{j})=0~~~\forall i,j.
\]

It can be shown that the necessary and sufficient condition for the reality of
a flat four-simplex geometry is that the scalar products of the bivectors
associated to the triangles be real \cite{ClassicalAreaReal}.

I\ would like to formally reduce the Barrett-Crane models for real general
relativity from that of the $SO(4,C)$ Barrett-Crane model by using the
bivector scalar products reality constraint \cite{ClassicalAreaReal}.
Precisely I\ plan to use the following three ideas to reduce the Barrett-Crane models:

\begin{enumerate}
\item The formal structure of the reduced intertwiners should be the same as
that of the $SO(4,C)$ Barrett-Crane model,

\item The eigen value of the Casimir corresponding to the square of the area
of any triangle must be real. I\ would like to refer to this as the
self-reality constraint\footnote{I\ would like to mention that the areas being
real necessarily does not mean that the bivectors must also be real.},

\item The eigen values of the square of area Casimir corresponding to the
representations associated with the internal links of the intertwiner must be
real. I\ would like to refer to this as the cross-reality constraint.
\end{enumerate}

The first idea sets a formal ansatz for the reduction process. The square of
the area of a triangle is simply the scalar product of the bivector of a
triangle with itself. Second condition is the quantum equivalent of the
reality of the scalar product of a bivector associated with a triangle with
itself. Once the second condition is imposed the third condition is the
quantum equivalent of the reality of the scalar product of the two bivectors
of any two triangle of a tetrahedron\footnote{We have ignored to impose
reality of the scalar products of the bivectors associated to any two
triangles of the same four simplex which intersect at only at one vertex. This
is because these constraints appears not to be needed for a formal extraction
of the Barrett-Crane models of real general relativity from that of $SO(4,C)$
general relativity described in this section. Imposing these constraints may
not be required because of the enormous redundancy in the bivector scalar
product reality constraints defined in Ref:\cite{ClassicalAreaReal}. This
issue need to be carefully investigated}.

My goal is to use the above principles to derive reduced Barrett-Crane models
and later one can convince oneself by identifying and verifying that the
Barrett-Crane constraints are satisfied for a subgroup of $SO(4,C)$ for each
of the reduced model.

In general by reducing a certain Hilbert space associated with the
representations of a group $G$ by some constraints, the resultant Hilbert
space need not contain the states gauge invariant under $G$. In that case one
can look for gauge invariance states under subgroups of $G$. In our case we
will find that the suitable quantum states extracted by adhering to the above
principles are gauge symmetry reduced versions of $SO(4,C)$ Barrett-Crane
states. They are gauge invariant only under the real subgroups of $SO(4,C)$.

Let $P$ be a formal projector which reduces the Hilbert space $D_{\chi_{L}%
}\otimes$ $D_{\chi_{R}}$ to a reduced Hilbert space such that the reality
constraints are satisfied. Let me assume as an ansatz that now the complex
three sphere is replaced by its subspace $X$ due to projection. Now I expect,
the projected $SO(4,C)$ Barrett-Crane intertwiner is spanned by the following
states for all $\chi_{i}$ satisfying the reality constraint in equation
(\ref{eq.rc}):%
\[
\Psi_{X}=\int_{x\in X}%
{\displaystyle\prod\limits_{i}}
PT_{\chi_{i}}(\mathfrak{g}(x))\tilde{d}\mathfrak{g}(x),
\]
where $\tilde{d}\mathfrak{g}(n)$ is the reduced measure of $d\mathfrak{g}(n)$
on $X.$ The imposition of the constraint expressed at the quantum level sets
$\rho_{i}$ or $n_{i}$ to be zero on each vertex of the $SO(4,C)$ Barrett-Crane
intertwiner. Let me rewrite the projected intertwiner as follows.%
\[
\Psi_{X}=\int_{x,y\in X}%
{\displaystyle\prod\limits_{1,2}}
PT_{\chi_{1}}(\mathfrak{g}(x))\delta_{X}(x,y)%
{\displaystyle\prod\limits_{3,4}}
PT_{\chi_{1}}(\mathfrak{g}(y))d^{X}g(x)d^{X}g(y),
\]
where $\delta_{X}(x,y)$ is the delta function on $X$. Since $X$ is a subspace
of $SL(2,C)$ a harmonic expansion can be derived for $\delta(x,y)$ using the
unitary representations of $SL(2,C)$. Since the intertwiner must obey the
cross reality constraint the harmonic expansion must only contain simple
representations of $SL(2,C)$ ($\rho$ or $n$ is zero).

For the Fourier transform defined by $PT_{\chi}(\mathfrak{g}(x))$ to be
complete and orthonormal I must have%
\[
\int_{\chi\in Q}\bar{\chi}\chi PT_{\chi}(\mathfrak{g}(x))PT_{\chi
}(\mathfrak{g}(y))d\chi=\delta_{X}(x,y),
\]
where $Q$ is the set of all simple representations\footnote{One could also
call the simple representations of $SL(2,C)$ as the real representations since
it corresponds to the real areas and the real homogenous spaces. But I\ will
avoid this to avoid any possible confusion.} of $SL(2,C)$ required for the
expansion. Only the simple representations of $SL(2,C)$ must be used to
satisfy the cross-reality constraint. Thus, the number of reduced intertwiners
derivable is directly related to the possible solutions for this equation
(subjected to Barrett-Crane constraints).

The equation of a complex three sphere is
\[
x^{2}+y^{2}+z^{2}+t^{2}=1.
\]
There are four different topologically different maximally connected real
subspaces of $CS^{3}$ such that the harmonic (Fourier) expansions on these
spaces use the simple representations of $SL(2,C)$ only. They are namely, the
three sphere $S^{3},$ the real hyperboloid $H^{+}$, the imaginary hyperboloid
$H^{-}$ and the Kleinien hyperboloid\footnote{By Kleinien hyperboloid I\ refer
to the space described by $x^{2}+y^{2}-z^{2}-t^{2}=1$ for real $x,y,z$ and
$t.$} $K^{3}$. Each of these subspace $X$ are maximal real subspaces of
$CS^{3}$. They are all homogenous under the action of a maximal real
subgroup\footnote{The real group is maximal in the sense that there is no
other real topologically connected subgroup of $SO(4,C)$ that is bigger.}
$G_{X}$ of $SO(4,C)$. There exists a $G_{X}$ invariant measure $d^{X}(x)$. The
reduced bivectors acting on the functions on $X$ effectively take values in
the Lie algebra of $G_{X}$. Since the measure $d^{X}(n)$ is invariant, the
reduced intertwiner is gauge invariant. So the intertwiner $\Psi_{X}$ must
correspond to the quantum general relativity for the group $G_{X}$.

Let the coordinates of $n=(x,y,z,t)$ be restricted to real values here after
in this section. Let me discuss the various reduced intertwiners:

\begin{enumerate}
\item $\rho=0$ case: This uses only the $\chi=(0,n)$ representations only.
This corresponds to $X=S^{3}$, satisfying%
\[
x^{2}+y^{2}+z^{2}+t^{2}=1,
\]
which is invariant under $SO(4,R)$. So this case corresponds to the Riemannian
general relativity. The appropriate projected $T-$functions are the
representation matrices of $SU(2)\approx S^{3}$ and the reduced measure is the
Haar measure of $SU(2).$ The intertwiner I get is the Barrett-Crane
intertwiner for the Riemannian general relativity. Here the $\chi^{\prime}s$
has been replaced by the $J^{\prime}s$ and the complex three sphere by the
real three sphere. The case of going from the $SO(4,C)$ Barrett-Crane model to
the Riemannian Barrett-Crane model is intuitive. It is a simple process of
going from complex three sphere to its subspace the real three sphere.

\item $n=0$ case: This uses $\chi=(\rho,0)$ representations only: This
corresponds to $X$ as a space-like hyperboloid (only one sheet) with
$G_{X}=SO(3,1,R)$:%
\[
x^{2}+y^{2}+z^{2}-t^{2}=1.
\]

The intertwiner now corresponds to the Lorentzian general relativity. This
intertwiner was introduced in $\cite{BCLorentzian}$. The unitary
representations of the Lorentz group on the real hyperboloid have been studied
by Gelfand and Naimarck \cite{IMG}, from which the $T-$functions are
\[
T_{\rho}(x)[\xi]=[\xi.x]^{\frac{1}{2}i\rho-1},
\]
where $\xi$ $\in$ null cone intersecting $t=1$ plane in the Minkowski space.
Here $\xi$ replaces $(z_{1},z_{2})$ in the $T-$function $T_{\chi}%
(\mathfrak{g}(x))(z_{1},z_{2})$of the $SO(4,C)$ Barrett-Crane Model. An
element $g$ $\in$ $SO(3,1)$ acts as a shift operator as follows:
\begin{subequations}
\label{eq.action}%
\begin{align*}
gT_{\rho}(x)[\xi]  &  =T_{\rho}(gx)[\xi]\\
&  =T_{\rho}(x)[g^{-1}\xi].
\end{align*}

This intertwiner was first introduced in $\cite{BCLorentzian}$.

\item Combination of $(0,n)\ $and$(\rho,0)$ representations: There are two
possible models corresponding to this case. One of them has $X\ $as the
Kleinien hyperboloid defined by%
\end{subequations}
\[
x^{2}+y^{2}-z^{2}-t^{2}=1,
\]

with $G_{X}=SO(2,2,R)$. Here the $X$ is isomorphic to $SU(1,1)\approx
SL(2,R)$. The intertwiner now corresponds to Kleinien general relativity (
$++--$ signature). The $T-$functions are of the form $T_{\chi}(\mathrm{k}%
(n))(z_{1},z_{2})$ where $z_{1}$ and $z_{2}$ takes real values only (please
refer to appendix $C$ ), $\chi\neq0$ and $\mathrm{k}$ is an isomorphism from
the Kleinien hyperboloid to $SU(1,1)$ defined by
\[
\mathrm{k}(n)=\left[
\begin{array}
[c]{cc}%
x-iy & z-it\\
z+it & x+iy
\end{array}
\right]  .
\]
The representations corresponding to the $n=0$ and $\rho=0$ cases are
qualitatively different. The representations corresponding to $\rho\neq0$ are
called the \textbf{continuous representations} and those to $n\neq0$ are
called the \textbf{discrete representations. }The action of $g\in SO(2,2,R)$
on the $T-$functions is
\[
gT_{\chi}(\mathrm{k}(x))=T_{\chi}(\mathrm{k}(g(x)),
\]
where $g(x)$ is the result of action of $g$ on $x\in X$.

\item The second model using both $(0,n)\ $and$(\rho,0)$ representations: This
corresponds to the time-like hyperboloid with $G_{X}=$ $SO(3,1)$,%
\[
x^{2}-y^{2}-z^{2}-t^{2}=1,
\]
where two vectors that differ just by a sign are identified as a single point
of the space $X$. The corresponding spin foam model has been introduced by
Barrett and Crane \cite{BCReimmanion}. It has been derived using a field
theory over group formalism by Rovelli and Perez \cite{RoPeModel}. Similar to
the previous case, I have both continuous and discrete representations, with
the $T-$functions given by%
\begin{align*}
T_{\rho}(x)[\xi]  &  =[\xi.x]^{\frac{1}{2}i\rho-1},\\
T_{n}(x)[l(a,\xi)]  &  =\exp(-2in\theta)\delta(a.\xi),
\end{align*}
where the $l(a,\xi)$ is an isotropic line\footnote{A line on an imaginary
hyperboloid \cite{IMG} is the intersection of a 2-plane of the Minkowski space
with it. The line is called isotropic if the Lorentzian distance between any
two points on it is zero. An isotropic line $l$ is described by the equation
$x=s\xi+x_{0},$ $x$ is the variable point on $l$, $x_{0}$ is any fixed point
on $l,$ and $\xi$ is a null-vector. For more information please refer to
\cite{IMG}} on the imaginary hyperboloid along direction $\xi$ going through a
point $a$ on the hyperboloid and the $\theta$ is the distance between
$l(a,\xi)$ and $l(x,\xi)$ given by $\cos\theta=a.x,$ where the dot is the
Lorentzian scalar product. I have for $g\in SO(3,1,R),$%
\begin{align*}
gT_{n}(x)[l(a,\xi)]  &  =T_{n}(x)[l(a,g\xi)]\\
&  =T_{n}(g^{-1}x)[l(a,\xi)],
\end{align*}
and the action of $g$ on continuous representations are defined similar to
equation (\ref{eq.action}). The corresponding spin foam model has been
introduced and investigated before by Rovelli and Perez \cite{RoPeModel}.
\end{enumerate}

\subsection{The Area Eigenvalues}

Using the $T-$functions described above, the intertwiners for real general
relativity can be constructed. Using these intertwiners, spin foam models
(Barrett-Crane) for the real general relativity theories of the various
different signatures can be constructed. The square of the area of a triangle
of a four-simplex for all signatures associated with a representation $\chi$
is described by the same formula\footnote{Please refer to the end of appendix
$C$ regarding the differences between the Casimers of $SL(2,C)$ and
\ $SU(1,1).$},%
\begin{align*}
\eta_{IK}\eta_{JL}\hat{B}^{IJ}\hat{B}^{KL}  &  =\left(  \chi^{2}-1\right)
\hat{I}\\
&  =\left(  \frac{n^{2}}{2}-\rho^{2}-1\right)  \hat{I},
\end{align*}
where only of $n$ and $\rho$ is non-zero. The square of the area is negative
or positive depending on whether $\rho$ or $n$ is non-zero. The negative
(positive) sign corresponds to a time-like (space-like) area.

\section{Further Considerations:}

\subsection{A Mixed Lorentzian Quantum Model.}

We have two intertwiners for the Lorentzian general relativity discussed in
the previous section, one corresponding to the space-like hyperboloid $H^{+}$
\cite{BCLorentzian} and another to the time hyperboloid $H^{-}$
\cite{BCLorentzian}, \cite{RoPeModel}. We can consider a tetrahedron to be
space-like (time-like) if it is associated with the intertwiner related to the
space-like (time-like) hyperboloid. This is justified because in the
semi-classical limit the tetrahedron becomes a space-like (time-like)
hypersurface \cite{SMSKn-1+1}, \cite{JWBCS}. I can construct quantum
amplitudes for a general four-simplex with each tetrahedron of the 4-simplex
either time-like or space-like. The intertwiners are straight forward to
construct. This model is a more general form of the Lorentzian Barrett-Crane
model. Let me next discuss the various propagators associated with this model.

The propagator from a space-like tetrahedron with an associated vector
$t_{1}\in H^{+}$ to another space-like tetrahedron in the same simplex with an
associated vector $t_{2}\in H^{+}$ is given by%
\begin{align*}
h_{\rho}^{++}(t_{1},t_{2})  &  =\int(t_{1}.l)^{-1+i\rho}(t_{2}.l)^{-1-i\rho
}dl\\
&  =\frac{4\pi\sin(\eta\rho)}{i\rho\cosh\eta},
\end{align*}
where the unit vector $l$ is an element of the positive light cone
intersecting $t=1$ hypersurface in the Minkowski space-time, $dl$ is the
measure on the intersection. This propagator has been introduced and discussed
by Barrett and Crane \cite{BCLorentzian}. The propagators between two
time-like tetrahedra were discussed by Rovelli-Perez. I refer the readers to
Ref:\cite{RoPeModel} for the details.

One can define a propagator between a space-like and a time-like tetrahedra
intersecting at a triangle associated with a continuous $(\rho\neq0)$
representation. The propagator from a time-like tetrahedron associated with a
vector $t\in H^{-}$ to a tetrahedron associated with a space-like vector $s\in
H^{+}$ is given by%
\[
h_{\rho}^{+-}(t,s)=\int(t.l)^{-1+i\rho}\left\vert s.l\right\vert ^{-1-i\rho
}dl,
\]
where the unit vector $l$ is the element of the positive light cone with time
component equal to $1$ and the $dl$ the measure on it. An important difference
between this propagator and the other two propagators discussed before is that
there is no completion relation for this propagator, such as
\[
\int_{\chi}h_{\chi}(x_{1},x_{2})d\chi=\delta_{X}(x_{1}-x_{2}),
\]
where a formal propagator between two elements $x_{1},x_{2}$ of some space $X$
is summed and integrated over all possible representations.

To calculate this integral, using the Lorentz invariance of the integral, I
can define the space time coordinates such that $t=(1,0,0,0),$ $s=(\sinh
\eta,0,0,\cosh\eta)$, $l=(1,n)$, where $n$ is a $3D$ unit vector expressed in
terms of $\theta,\phi$ coordinates. Then the integral is%
\begin{align*}
h_{\rho}(t,s)  &  =\int(t.l)^{-1+i\rho}\left\vert s.l\right\vert ^{-1-i\rho
}dl\\
&  =\int\left\vert \left(  \sinh\eta-\cos\theta\cosh\eta\right)  \right\vert
^{-1-i\rho}\sin\theta d\theta d\phi\\
&  =-2\pi\int_{-1}^{+1}\left\vert \sinh\eta-z\cosh\eta\right\vert ^{-1-i\rho
}dz,
\end{align*}
where the $\cos\theta$ has been replaced by a new variable $z.$ Let
$q=\sinh\eta-z\cosh\eta$. When $z$ varies between $-1$ and $+1$, $q$ varies
between $e^{\eta}$ and $-e^{-\eta}$. In this range $q$ is zero only once when
$z=\tanh(\eta)$. Rewriting the above integral using $q$ as the variable of
integration I get,%
\begin{align*}
h_{\rho}(t,s)  &  =\frac{-2\pi}{\cosh\eta}\left\{  \int_{0}^{e^{\eta}%
}q^{-1-i\rho}dq+\int_{-e^{-\eta}}^{0}\left(  -q\right)  ^{-1-i\rho}dq\right\}
\\
&  =\frac{-2\pi}{\cosh\eta}\left\{  \int_{0}^{e^{\eta}}q^{-1-i\rho}dq+\int
_{0}^{e^{-\eta}}q^{-1-i\rho}dq\right\}  .
\end{align*}
By setting $q=e^{x},$ I get%
\[
h_{\rho}(t,s)=\frac{-2\pi}{\cosh\eta}\left\{  \int_{-\infty}^{\eta}e^{-i\rho
x}dx+\int_{-\infty}^{-\eta}e^{-i\rho x}dx\right\}  .
\]
This integral does not have a clear limit. But by assuming that $\rho$ has a
small positive imaginary part I get the following result\footnote{It is not
clear what is the physical meaning of the small positive imaginary part is.}:%
\[
h_{\rho}(t,s)=\frac{-2\pi i\cos[\rho\eta]}{\rho\cosh\eta}.
\]

At this point it is also not clear whether the new model has any physical
significance. Further investigation is required.

\subsection{A Multi-Signature Barrett-Crane Model}

I formally deduced various intertwiners corresponding to the various
signatures of real general relativity from the $SO(4,C)$ Barrett-Crane
intertwiner. By using each of these intertwiners I can construct a quantum
four-simplex for each signature. By splicing the quantum four simplices of the
various signatures on the tetrahedrons with common representations I can
construct a spin foam model. This model could be considered as the most
general Barrett-Crane model for real general relativity. The physical
significance of this model is not clear and further study is required.

Putting together quantum general relativity models of various signatures has
been considered before. For example, Hawking \cite{HawkingPenrose} has spliced
together a Euclidean geometry (imaginary time) universe in the initial stage
of the universe to its Lorentzian future. But the Hawking theory is slightly
different from mine. In the Hawking's theory the Euclidean general relativity
has an imaginary action and so it contributes magnitudes instead of phases to
the path integral. In our theory the action that is used for the spin foam
quantization is always real as is described in Ref:\cite{ClassicalAreaReal}.

It has been suggested before that for quantum general relativity to be unitary
it must involve all the signatures \cite{UnitarySignature}. So the classical
and quantum multi-signature real general relativity may be interesting new
theories to look into and explore for new physics.

\subsection{The Asymptotic Limit of the Barrett-Crane models.}

The asymptotic limit of the real Barrett-Crane models has been studied before
\cite{JWBCS}, \cite{JWBRW}, \cite{PonzanoReggeModel}, \cite{BaezEtalAsym} to a
certain degree. Here I will discuss the asymptotic limit of the $SO(4,C)$
Barrett-Crane model. For the first time I show here that we can extract
bivectors which satisfy the essential Barrett-Crane constraints from the
asymptotic limit. Consider the amplitude of a four-simplex given by Eq.
(\ref{10w}) with a real scale parameter $\lambda$,%
\begin{align*}
Z_{\lambda}  &  =\int_{n_{k}\in CS^{3}}%
{\displaystyle\prod\limits_{i<j}}
G_{\lambda\chi_{ij}}(n_{i}\mathfrak{,}n_{j})%
{\displaystyle\prod\limits_{k}}
dn_{k},\\
&  =\int_{n_{k}\in CS^{3}}%
{\displaystyle\prod\limits_{i<j}}
\dfrac{\cos(\lambda\rho_{ij}\eta_{ij}+\lambda n_{ij}\theta_{ij})}{\left\vert
\sinh(\lambda\eta_{ij}+i\lambda\theta_{ij})\right\vert ^{2}}%
{\displaystyle\prod\limits_{k}}
dx_{k},\\
&  =\int_{n_{k}\in CS^{3}}%
{\displaystyle\prod\limits_{i<j}}
\sum\limits_{\varepsilon_{ij}=\pm1}\dfrac{\exp(i\varepsilon_{ij}\lambda
(\rho_{ij}\eta_{ij}+n_{ij}\theta_{ij}))}{2\left\vert \sinh(\lambda\eta
_{ij}+i\lambda\theta_{ij})\right\vert ^{2}}%
{\displaystyle\prod\limits_{k}}
dx_{k},
\end{align*}
where the $\eta_{ij}+i\theta_{ij}$ is defined by $n_{i}.n_{j}=\cosh(\eta
_{ij}+i\theta_{ij})$. Here the $\zeta_{ij}=$ $\eta_{ij}+i\theta_{ij}$ is the
complex angle between $n_{i}$ and $n_{j}$. The asymptotic limit of
$Z_{\lambda}(s)\ $under $\lambda\longrightarrow\infty$ is controlled by%

\begin{align*}
&  S(\{n_{i},\bar{n}_{i}\},\{\chi_{ij},\bar{\chi}_{ij}\})\\
&  =\sum_{i<j}\varepsilon_{ij}(\rho_{ij}\eta_{ij}+n_{ij}\theta_{ij}%
)+\operatorname{Re}\left(  \sum_{i}q_{i}(n_{i}.n_{i}-1)\right) \\
&  =\operatorname{Re}\left(  \sum_{i<j}\varepsilon_{ij}\bar{\chi}_{ij}%
\zeta_{ij}+\sum_{i}q_{i}(n_{i}.n_{i}-1)\right)  ,
\end{align*}
where the $q_{i}$ are the Lagrange multipliers to impose $n_{i}.n_{i}%
=1,\forall i$. My goal now is to find stationary points for this action. The
stationary points are determined by
\begin{subequations}
\begin{equation}
\sum_{~~~~~i\neq j}\varepsilon_{ij}\bar{\chi}_{ij}\frac{\partial\zeta_{ij}%
}{\partial n_{i}}+q_{j}n_{j}=0,\forall j, \label{ext}%
\end{equation}
and $n_{j}.n_{j}=1,\forall j$ where the $j$ is a constant in the summation.%

\end{subequations}
\begin{equation}
\frac{\partial\zeta_{ij}}{\partial n_{i}}=\frac{n_{j}}{\sinh(\zeta_{ij})}.
\label{eq.dif.as}%
\end{equation}

Using equation (\ref{eq.dif.as}) in equation :(\ref{ext}) and taking the wedge
product of the equation with $n_{j}$ we have,
\[
\sum_{~~~~~i\neq j}\varepsilon_{ij}\bar{\chi}_{ij}\frac{n_{j}\wedge n_{j}%
}{\sinh(\zeta_{ij})}=0,\forall j.
\]

If
\[
\bar{E}_{ij}=i\varepsilon_{ij}\bar{\chi}_{ij}\frac{n_{j}\wedge n_{j}}%
{\sinh(\zeta_{ij})},
\]
then the last equation can be simplified to%

\begin{equation}
\sum_{~~~~~i\neq j}E_{ij}=0,\forall j. \label{sumzero}%
\end{equation}

We now consider the properties of $E_{ij}:$

\begin{itemize}
\item Each $i$ represents a tetrahedron. There are ten $E_{ij}$'s, each one of
them is associated with one triangle of the four-simplex.

\item The square of $E_{ij}$:%
\begin{align*}
\bar{E}_{ij}\cdot\bar{E}_{ij}  &  =\frac{-\bar{\chi}_{ij}^{2}}{\sinh^{2}%
(\zeta_{ij})}(n_{j}^{2}n_{i}^{2}-\left(  n_{i}\cdot n_{j}\right)  ^{2})\\
&  =\frac{-\bar{\chi}_{ij}^{2}}{\sinh^{2}(\zeta_{ij})}(1-\left(  \cosh
(\zeta_{ij}\right)  ^{2})\\
&  =\bar{\chi}_{ij}^{2}.
\end{align*}

\item The wedge product of any two $E_{ij}$ is zero if they are equal to each
other or if their corresponding triangles belong to the same tetrahedron.

\item Sum of all the $E_{ij}$ belonging to the same tetrahedron are zero
according to equation (\ref{sumzero}).
\end{itemize}

It is clear that these properties contain the first four Barrett-Crane
constraints. So we have successfully extracted the bivectors corresponding to
the triangles of a general flat four-simplex in $SO(4,C)$ general relativity
and the $n_{i}$ are the normal vectors of the tetrahedra. The $\chi_{ij}$ are
the complex areas of the triangle as one would expect. Since we did not impose
any non-degeneracy conditions, it is not guaranteed that the tetrahedra or the
four-simplex have non-zero volumes.

The asymptotic limit of the partition function of the entire simplicial
manifold with triangulation $\Delta$ is%
\[
S(\Delta,\{n_{is}\in CS^{3},\chi_{ij},\bar{\chi}_{ij},\varepsilon
_{ijs}\})=\operatorname{Re}\sum_{i<j,s}\varepsilon_{ijs}\bar{\chi}_{ij}%
\zeta_{ijs},
\]
where I\ have assumed variable $s$ represents the four simplices of $\Delta$
and $i,$ $j$ represents the tetrahedra. The $\varepsilon_{ijs}$ can be
interpreted as the orientation of the triangles. Each triangle has a
corresponding $\chi_{ij}$. The $n_{is}$ denote the unit complex vector
associated with the side of the tetrahedron $i$ facing the inside of a simplex
$s$. Now there is one bivector $E_{sij}$ associated with each side facing
inside of a simplex $s$ of a triangle $ij$ defined by%

\[
\bar{E}_{ijs}=i\varepsilon_{ijs}\bar{\chi}_{ij}\frac{n_{js}\wedge n_{js}%
}{\sinh(\zeta_{ijs})}.
\]
If the $n_{is}$ are chosen such that they satisfy stationary conditions%

\[
\sum_{~~~~~i\neq j}E_{ijs}=0,\forall j,s,
\]
and if
\[
\theta_{ij}=\left(  \sum_{s}\varepsilon_{ijs}\zeta_{ijs}\right)  ,
\]

then%

\begin{align*}
S(\Delta,\{\chi_{ij},\bar{\chi}_{ij},\varepsilon_{ijs}\})  &
=\operatorname{Re}\sum_{i<j,s}\varepsilon_{ijs}\bar{\chi}_{ij}\zeta_{ijs},\\
&  =\operatorname{Re}\sum_{i<j}\bar{\chi}_{ij}\theta_{ij}%
\end{align*}
can be considered to describe the Regge calculus for the $SO(4,C)$ general
relativity. The angle $\theta_{ij}$ are the deficit angles associated with the
triangles and the $n_{is}$ are the complex vector normals associated with the
tetrahedra. From the analysis that has been done in this section, it is easy
see that the $SO(4,C)$ Regge calculus contains the Regge calculus theories for
all the signatures. The Regge calculus for each signature can be obtained by
restricting the $n_{is}$ and the $\chi_{ij}$ to the corresponding homogenous
space and representations as described in the previous section. Also by the
properly restricting the $n_{is}$ and the $\chi_{ij}$ we can derive the Regge
calculus corresponding to the mixed Lorentzian and multi-signature
Barrett-Crane models described in the previous subsections.

\subsection{3+1 Formulation: Spin Networks Functionals.}

\begin{figure}
[ptbh]
\begin{center}
\includegraphics[
height=3.3053in,
width=3.8744in
]%
{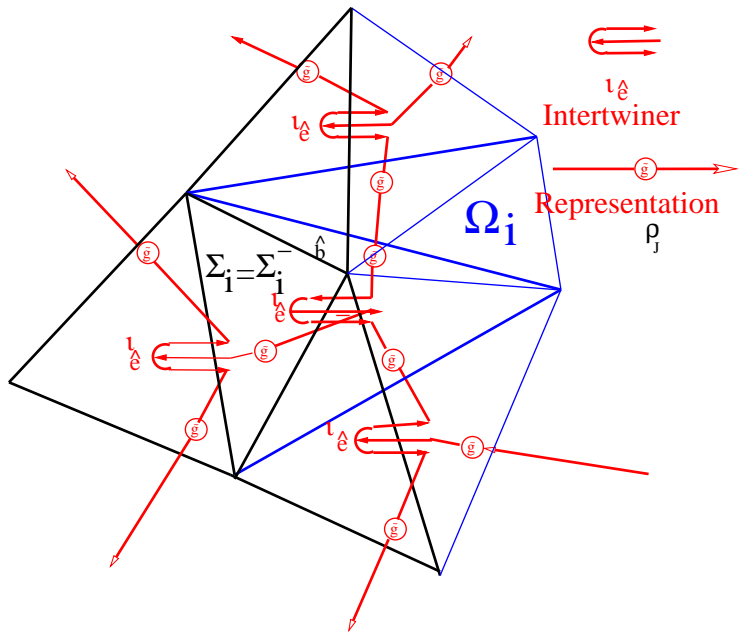}%
\caption{In this diagram $\Sigma_{i}$ is the $i^{th}$ hypersurface of the
foliation. $\Omega_{i}$ is the four dimensional slice between $\Sigma_{i}$ and
$\Sigma_{i+1}$. $\Sigma_{i+1}$ is not shown. $\hat{b}$ and $\hat{e}$ are the
triangles and tetrahedra on $\Sigma_{i}$. The links that go between the
hypersurfaces are shown in blue. The $\tilde{g}$'s are the parallel
propagators associated to the tetrahedra that go between the hypersurfaces. }%
\label{fig.spin}%
\end{center}
\end{figure}

A $(n-1)+1$ formulation was proposed in Ref:\cite{SMSKn-1+1}, with a
motivation to relate spin foams to canonical quantum general relativity.
I\ will briefly review the basic ideas and discuss it in the context of the
$SO(4,C)$ general relativity. For details, I refer to the original article
Ref:\cite{SMSKn-1+1}. The $nD$ simplicial manifold was foliated by a one
parameter sequence of $(n-1)D$ simplicial hypersurfaces. The parallel
propagators associated with the edges in the foliating hypersurfaces can be
thought of as the analog of the continuum connection in the (coordinate) time
direction. It turns out that the integration of the Feynman weight $e^{iS}$
with respect to the parallel propagators associated with the edges of the
hypersurfaces results in a product of spin network functionals shown in figure
(\ref{fig.spin}). These spin network functionals are defined on the parallel
propagators associated with the edges that go between the hypersurfaces on the
graphs that are dual to the triangulation of the foliating hypersurfaces. For
the spin foam model of the BF theory, the BF intertwiners are used to
intertwine the representations associated to the links of the graph. In the
case of spin foam model of general relativity, the Barrett-Crane intertwiners
are used. In this case the elements of the homogenous space on which the
intertwiners are defined represent normal vectors to the simplicial
hypersurfaces. The sum over the homogenous space vectors in the Barrett-Crane
intertwiners can be interpreted as a sum over the normals. The spin foam
partition functionals of the BF theories or general relativity (Barrett-Crane)
can be reformulated using these spin network functionals.%

It is straight forward to generalize the 3+1 theory to the $SO(4,C)$ general
relativity. By using the Barrett-Crane intertwiners for $SO(4,C)$ and the
$SO(4,C)$ parallel propagators, we can reconstruct the spin network
functionals for the $SO(4,C)$ general relativity. In this case the normal
vectors are complex. The $SO(4,C)$ 3+1~formulation essentially contains the
$3+1$ formulation of the real general relativity theories. It is also
straightforward to see that spin network functional for the real general
relativity models are: 1) restrictions of the complex normal vectors to the
real normal vectors and 2) restrictions of representations as described in
section three. These ideas are also applicable to the mixed Lorentzian and
multi-signature Barrett-Crane models by using the appropriate intertwiners,
representations and the parallel propagators.

\subsection{Field Theory over Group and Homogenous Spaces.}

One of the problems with the Barrett-Crane model for general relativity is its
dependence on the discretization of the manifold. A discretization independent
model can be defined by summing over all possible discretizations. With a
proper choice of amplitudes for the lower dimensional simplices the BF\ spin
foams can be reformulated as a field theory over a group (GFT)
\cite{ooguriBFderv}. Similarly, the Barrett-Crane models can be reformulated
as a field theory over the homogenous space of the group \cite{GFTreimm}. Let
me briefly explain the GFT of the four dimensional spin foams for a compact
group $G$. A\ field theory over a group is defined using an action. The action
has two terms, namely the kinetic term and the potential terms. Consider a
tetrahedron. Let a group element $g_{i}$ be associated with each triangle $i$
of the tetrahedron. Let a real field $\phi(g_{1},g_{2},$ $g_{3},g_{4})$
invariant under the exchange of its arguments be associated with the
tetrahedron. Let the field be invariant under the simultaneous (left or right)
action of a group element $g$ on its variables. Then the kinetic term is
defined as%

\[
K.E=\int\prod\limits_{i=1}^{4}dg_{i}\phi^{2}.
\]

To define the potential term, consider a four-simplex. Let $g_{i},$ where
$i=1$ to $10$ be the group elements associated with its ten triangles. With
each tetrahedron $e$ of the four-simplex, associate a $\phi$ field which is a
function of the group elements associated with its triangles. Denote it as
$\phi_{e}$. Then the potential term is defined as%
\begin{align*}
P.E  &  =\\
&  =\frac{\lambda}{5!}\int\prod\limits_{i=1}^{10}dg_{i}\prod\limits_{e=1}%
^{5}\phi_{e},
\end{align*}
where $\lambda$ is an arbitrary constant. Now the action for a GFT can be
defined as%
\[
S(\phi)=K.E+P.E=\int\prod\limits_{i=1}^{4}dg_{i}\phi^{2}+\frac{\lambda}%
{5!}\int\prod\limits_{i=1}^{10}dg_{i}\prod\limits_{e=1}^{5}\phi_{e}.
\]
The Partition function of the $GFT$ is%

\[
Z=\int D\phi e^{-S(\phi)}.
\]
Now, an analysis of this partition function yields the sum over spin foam
partitions of the four dimensional BF theory for group $G$ for all possible
triangulations. From the analysis of the GFT we can easily show that this
result is valid for $G=$ $SO(4,C)$ with the unitary representations defined in
the appendix B.

Let us assume $\phi$ is invariant only under the simultaneous action of an
element of a subgroup $H$ of $G$. Then, if $G=SO(4,R)$ and $H=SU(2)$ we get
GFTs for the Barrett-Crane model\footnote{Depending on whether we are using
the left or right action of $G\ $on $\phi$, we get two different models that
differ by amplitudes for the lower dimensional simplices \cite{GFTreimm}.}
\cite{GFTreimm}. Similarly, if $G=$ $SL(2,C)$ and $H=SU(2)$ or $SU(1,1)$, we
can define GFT for the Lorentzian general relativity \cite{RoPeModel},
\cite{RovPerGFTLorentz}. The representation theories of $SO(4,C)$ and
$SL(2,C)$ has similar structure to those of $SO(4,R)$ and $SU(2)$
respectively. So the GFT\ with $G=SO(4,C)$ and $H=SL(2,C)$ should yield the
sum over triangulation formulation of the $SO(4,C)\ $Barrett-Crane model. The
details of this analysis and its variations will be presented elsewhere.

\section{Summary}

In this article I\ have comprehensively investigated various issues involved
in the formulation of the spin foam models for general relativity. In this
process many things has been accomplished. They can be listed as follows:

\begin{itemize}
\item Formulated the spin foams for the $SO(4,C)\ $BF theory.

\item Systematically imposed the essential Barrett-Crane constraints for the
$SO(4,C)$ general relativity:

\begin{itemize}
\item Rigorously imposed the cross-simplicity constraints for the $SO(4,C)$
general relativity. This procedure can be directly applied to the Riemannian
general relativity.

\item The Barrett-Crane intertwiner for the $SO(4,C)$ general relativity has
been calculated

\item The propagators of the $SO(4,C)$ Barrett-Crane model has been calculated
and the four-simplex amplitude was formulated using them.
\end{itemize}

\item Using the bivector scalar product reality condition the Barrett-Crane
intertwiners for all non-degenerate signatures have been formally deduced.

\item Discussed the asymptotic limit the $SO(4,C)$ general relativity which
can be easily restricted to the real general relativity cases. Essentially the
asymptotic limit is the $SO(4,C)$ Regge calculus which contains the Regge
calculus theories of all the real general relativity cases.

\item The $3+1$ formulations of the $SO(4,C)\ $and the real Barrett-Crane
models have been briefly discussed.

\item Field theory over group for the $SO(4,C)$ Barrett-Crane model has been
briefly introduced.

\item We discussed the mixed Barrett-Crane models which mixes the intertwiners
for the two Lorentzian Barrett-Crane models and calculate the mixed propagator.

\item Proposed a multi-signature Barrett-Crane model which is obtained by
coupling together the four-simplex amplitudes for the various different signatures.
\end{itemize}

\subsection{Future directions}

All of the above listed work focused on defining, deriving, and achieving
general and unified understanding of the spin foam models of general
relativity. This process is still incomplete. Also the process of extracting
physics from the spin foams is incomplete and little explored. Let me list the
future work that needs to be done in successfully formulating a coordinate and
back-ground independent quantum general relativity theory.

\begin{itemize}
\item Completing the Model:

\begin{itemize}
\item Fixing the degenerate contributions: A\ careful study of the asymptotic
limit for the Riemannian Barrett-Crane model \cite{JWBCS} revealed the
existence of the degenerate contributions. These contributions are not only
present in the model but they also dominate the asymptotic limit. This could
be considered as a result of not having a nice way to impose the last two
Barrett-Crane constraints which ensure the non-zero volumes. It is possible
that the semi-classical limit may not be the same as the asymptotic limit. If
the semiclassical limit is related to the asymptotic limit then one must find
a physical explanation for the degenerate contributions or find a way to fix
them. The semiclassical limit issues will be further discussed below.

\item Sum over the spin foams and unitarity: Spin foam models are essentially
the path integral quantization of the discretized gravitational actions. The
spin foam amplitudes are the quantum transition amplitudes between spin
networks. The spin foam transition amplitudes from one spin network associated
with a hypersurface to another spin network associated with a similar
hypersurface can be calculated. Assume these two spin networks are associated
with graphs (dual triangulations) of different sizes. Then the Hilbert space
associated with the two hypersurfaces need not be of the same size. In this
case the spin foam transition amplitudes are clearly not unitary. Thus it
becomes necessary that a sum over triangulation may need to be performed to
realize unitarity, which directly leads to the field theory over group space
formulations. But after performing the summation it is not clear and it is not
yet shown that the spin foam transition amplitudes are unitary. So, this issue
needs to be explored.
\end{itemize}

\item Extracting Physics: An important area that needs to be explored is to
understand the relation between classical physics and spin foam models.
Clearly the spin foams are founded on the discretization and the quantization
ideas of classical general relativity. But, extracting the classical physics
from spin foam models needs to be done at various levels.

\begin{itemize}
\item Semiclassical and continuum limit issues: An important question is how
to calculate rigorously the semiclassical limit for quantum general
relativity. The calculation of the semiclassical limits is a less understood
problem even in the conventional quantum mechanics itself \cite{SemiClasQuest}%
. In the case of spin foam models the asymptotic limit is usually considered
to be the semi-classical limit of quantum general relativity. This has been
motivated by the idea of semi-classical limits of the angular momentum
\cite{AngMomClassLimit}. But even though the mathematics involved in the spin
foams is that of angular momentum calculus, the physics is not the same. So,
the question is whether the asymptotic limit is the same as the semiclassical
limit? If not, what is the semiclassical limit?

\item Convergence issues and amplitudes of lower dimensional simplices: The
convergence issues involved in spin foam models are different from that
involved in the usual perturbative quantum field theories like QED. In
perturbative QED we are summing a series of terms (Feynman diagrams)
corresponding to the various orders of the coupling constant. Calculating more
elements of the series increases the precision of the quantity calculated. For
the case of spin foam models this interpretation clearly does not hold. An
important question is whether the partition functions of the spin foams are
convergent? In general the partition functions spin foams do not converge
\cite{PerezReview}. Modified Barrett-Crane models for real general relativity
based on GFTs were proposed \cite{RovPerNoDiv}. It has been demonstrated that
by choosing proper choices of quantum amplitudes for simplices in less than
three dimensions a convergent spin foam model can be defined
\cite{RovPerNoDiv}. Before investigating convergence issues, the next
important step might be the construction of a spin foam model with a
physically motivated choice of amplitudes for lower dimensional simplices
addressing all the issues discussed before and any other relevant issues.
\end{itemize}

\item Unification with Matter: Unification with matter needs to investigated
for the $SO(4,C)$ general relativity with the reality constraint for both the
quantum and classical case. Currently there are various proposals and studies
for the inclusion matter in case of the four dimensional and the three
dimensional general relativity theories \cite{Matter}.

\item Relationship with Canonical Quantum general relativity:

\begin{itemize}
\item Relationship to canonical quantum general relativity needs to be
investigated. My work on the $(n-1)+1$ formulation of the spin foams
\cite{SMSKn-1+1} explicitly demonstrates how to relate the spin network
functionals of canonical quantum general relativity to the spin foams of BF
theory and General Relativity. But the important issues are in interpreting
the diffeomorphism and Hamiltonian constraints in the context of the spin foams.

\item An important question that arises is whether there is a rigorous
relationship of the reality constraint in the spin foams to that of the
reality condition \cite{AABOOK} in canonical quantum general relativity?
Realization of the reality condition quantum mechanically is a non-trivial
problem in canonical quantum general relativity. In fact many of the recent
advances \cite{MB} in canonical quantum general relativity have been made by
converting the complex formulation of the theory to a real formulation by
transforming the configuration variable a complex $SL(2,C)$ connection to a
real $SU(2)$ connection through a Legendre transformation \cite{JFB}.
\end{itemize}
\end{itemize}

\section{Acknowledgement.}

I thank Allen Janis, George Sparling and John Baez for correspondences.

\appendix

\section{Unitary Representations of SL(2,$\boldsymbol{C}$)}

The Representation theory of $SL(2,\boldsymbol{C})$ was developed by Gelfand
and Naimarck \cite{IMG}. Representation theory of $SL(2,C)$ can be developed
using functions on $C^{2}$ which are homogenous in their
arguments\footnote{These functions need not be holomorphic but infinitely
differentiable may be except at the origin $(0,0)$.}. The space of functions
$D_{\chi}$ is defined as functions $f(z_{1},z_{2})$ on $C^{2}$ whose
homogeneity is described by%
\[
f(az_{1},az_{2})=a^{\chi_{1}-1}a^{\chi_{2}-1}f(z_{1},z_{2}),
\]
for all $a\neq0,$ where $\chi$ is a pair $(\chi_{1},\chi_{2})$. The linear
action of $SL(2,C)$ on $C^{2}$ defines a representation of $SL(2,C)$ denoted
by $T_{\chi}$. Because of the homogeneity of functions of $D_{\chi},$ the
representations $T_{\chi}$ can be defined by its action on the functions
$\phi(z)$ of one complex variable related to $f(z_{1},z_{2})\in$ $D_{\chi}$ by%
\[
\phi(z)=f(z,1).
\]
There are two qualitatively different unitary representations of $SL(2,C)$:
the principal series and the supplementary series, of which only the first one
is relevant to quantum general relativity. The principal unitary irreducible
representations of $SL(2,\boldsymbol{C})$ are the infinite dimensional. For
these $\chi_{1}=-\bar{\chi}_{2}=\frac{n+i\rho}{2},$ where $n$ is an integer
and $\rho$ is a real number. In this article I\ would like to label the
representations by a single complex number $\chi=\frac{n}{2}+i\frac{\rho}{2}$,
wherever necessary. The $T_{\chi}$ representations are equivalent to
$T_{-\chi}$ representations \cite{IMG}.

Let $g$ be an element of $SL(2,\boldsymbol{C})$ given by%
\[
g=\left[
\begin{array}
[c]{cc}%
\alpha & \beta\\
\gamma & \delta
\end{array}
\right]  ,
\]
where $\alpha$,$\beta$,$\gamma$ and $\delta$ are complex numbers such that
$\alpha\delta-\beta\delta=1$. Then the $D\chi$ representations are described
by the action of a unitary operator $T_{\chi}(g)$ on the square integrable
functions $\phi(z)$ of a complex variable $z$ as given below:%
\begin{equation}
T_{\chi}(g)\phi(z)=(\beta z_{1}+\delta)^{\chi-1}(\bar{\beta}\bar{z}_{1}%
+\bar{\delta})^{-\bar{\chi}-1}\phi(\frac{\alpha z+\gamma}{\beta z+\delta}).
\label{rep}%
\end{equation}
This action on $\phi(z)$ is unitary under the inner product defined by%
\[
\left(  \phi(z),\eta(z)\right)  =\int\bar{\phi}(z)\eta(z)d^{2}z,
\]
where $d^{2}z=\frac{i}{2}dz\wedge d\bar{z}$ and I\ would like to adopt this
convention everywhere. Completing $D_{\chi}$ with the norm defined by the
inner product makes it into a Hilbert space $H_{\chi}$.

Equation (\ref{rep}) can also be written in kernel form
\cite{RovPerGFTLorentz},%
\[
T_{\chi}(g)\phi(z_{1})=\int T_{\chi}(g)(z_{1},z_{2})\phi(z_{2})d^{2}z_{2},
\]
Here $T_{\chi}(g)(z_{1},z_{2})$ is defined as%
\begin{equation}
T_{\chi}(g)(z_{1},z_{2})=(\beta z_{1}+\delta)^{\chi-1}(\bar{\beta}\bar{z}%
_{1}+\bar{\delta})^{-\bar{\chi}-1}\delta(z_{2}-g(z_{1})), \label{eq.rep}%
\end{equation}
where $g(z_{1})=\frac{\alpha z_{1}+\gamma}{\beta z_{1}+\delta}$. The Kernel
$T_{\chi}(g)(z_{1},z_{2})$ is the analog of the matrix representation of the
finite dimensional unitary representations of compact groups. An infinitesimal
group element, $a$, of $SL(2,\boldsymbol{C})$ can be parameterized by six real
numbers $\varepsilon_{k}$ and $\eta_{k}$ as follows \cite{Ruhl}:%
\[
a\approx I+\frac{i}{2}\sum_{k=1}^{3}(\varepsilon_{k}\sigma_{k}+\eta_{k}%
i\sigma_{k}),
\]
where the $\sigma_{k}$ are the Pauli matrices. The corresponding six
generators of the $\chi$ representations are the $H_{k}$ and the $F_{k}$. The
$H_{k}$ correspond to rotations and the $F_{k}$ correspond to boosts. The
bi-invariant measure on $SL(2,C)$ is given by
\[
dg=\left(  \frac{i}{2}\right)  ^{3}\frac{d^{2}\beta d^{2}\gamma d^{2}\delta
}{\left\vert \delta\right\vert ^{2}}=\left(  \frac{i}{2}\right)  ^{3}%
\frac{d^{2}\alpha d^{2}\beta d^{2}\gamma}{\left\vert \alpha\right\vert ^{2}}.
\]
This measure is also invariant under inversion in $SL(2,\boldsymbol{C})$. The
Casimir operators for $SL(2,C$ $)$ are given by%
\[
\hat{C}=\det\left[
\begin{array}
[c]{cc}%
\hat{X}_{3} & \hat{X}_{1}-i\hat{X}_{2}\\
\hat{X}_{1}+i\hat{X}_{2} & -\hat{X}_{3}%
\end{array}
\right]
\]
and its complex conjugate $\bar{C}$ where $X_{i}=F_{i}+iH_{i}.$ The action of
$C$ ($\bar{C}$) on the elements of $D_{\chi}$ reduces to multiplication by
$\chi_{1}^{2}-1$ ($\chi_{2}^{2}-1$).The real and imaginary parts of $C$ are
another way of writing the Casimirs. On $D_{\chi}$ they reduce to the
following%
\begin{align*}
\operatorname{Re}(\hat{C})  &  =\left(  -\rho^{2}+\frac{n}{4}^{2}-1\right)
\hat{I},\\
\operatorname{Im}(\hat{C})  &  =\rho n\hat{I}.
\end{align*}

The Fourier transform theory on $SL(2,\boldsymbol{C})$ was developed in
Ref:\cite{IMG}. If $f(g)$ is a square integrable function on the group, it has
a group Fourier transform defined by%
\begin{equation}
F(\chi)=\int f(g)T_{\chi}(g)dg, \label{Four}%
\end{equation}
where is $F(\chi)$ is linear operator defined by the kernel $K_{\chi}%
(z_{1},z_{2})$ as follows:%
\[
F(\chi)\phi(z)=\int K_{\chi}(z,\acute{z})\phi(\acute{z})d^{2}\acute{z}.
\]
The associated inverse Fourier transform is%
\begin{equation}
f(g)=\frac{1}{8\pi^{4}}\int Tr(F(\chi)T_{\chi}(g^{-1}))\chi\bar{\chi}d\chi,
\label{invFour}%
\end{equation}
where the $\int d\chi$ indicates the integration over $\rho$ and the summation
over $n.$ From the expressions for the Fourier transforms, I can derive the
orthonormality property of the $T_{\chi}$ representations,%
\[
\int_{SL(2,C)}T_{\acute{z}_{1}\chi_{1}}^{z_{1}}(g)T_{\acute{z}_{2}\chi_{2}%
}^{\dag z_{2}}(g)dg=\frac{8\pi^{4}}{\chi_{1}\bar{\chi}_{1}}\delta(\chi
_{1}-\chi_{2})\delta(z_{1}-\acute{z}_{1})\delta(z_{2}-\acute{z}_{2}),
\]
where $T_{\chi}^{\dagger}$ is the Hermitian conjugate of $T_{\chi}$.

The Fourier analysis on $SL(2,C)$ can be used to study the Fourier analysis on
the complex three sphere $CS^{3}$. If $x=(a,b,c,d)\in$ $CS^{3}$ then the
isomorphism $\mathfrak{g}:CS^{3}\longrightarrow SL(2,C)\ $can be defined by
the following:%
\[
\mathfrak{g}(x)=\left[
\begin{array}
[c]{cc}%
a+ib & c+id\\
-c+id & a-ib
\end{array}
\right]  .
\]
Then, the Fourier expansion of $f(x)$ $\in L^{2}(CS^{3})$ is given by%
\[
f(x)=\frac{1}{8\pi^{4}}\int Tr(F(\chi)T_{\chi}(\text{$\mathfrak{g}$}%
(x)^{-1})\chi\bar{\chi}d\chi
\]
and its inverse is
\[
F(\chi)=\int f(g)T_{\chi}(\mathfrak{g}(x))dx,
\]
where the $dx$ is the measure on $CS^{3}$. The measure $dx$ is equal to the
bi-invariant measure on $SL(2,C)$ under the isomorphism $\mathfrak{g}$.

The expansion of the delta function on $SL(2,C)$ from equation (\ref{invFour})
is%
\begin{equation}
\delta(g)=\frac{1}{8\pi^{4}}\int tr\left[  T_{\chi}(g)\right]  \chi\bar{\chi
}d\chi. \label{deltaExp}%
\end{equation}
Let me calculate the trace $tr\left[  T_{\chi}(g)\right]  $. If $\lambda
=e^{\rho+i\theta}$ and $\frac{1}{\lambda}$ are the eigen values of $g$ then%
\[
tr\left[  T_{\chi}(g)\right]  =\dfrac{\lambda^{\chi_{1}}\bar{\lambda}%
^{\chi_{2}}+\lambda^{-\chi_{1}}\bar{\lambda}^{-\chi_{2}}}{\left\vert
\lambda-\lambda^{-1}\right\vert ^{2}},
\]
which is to be understood in the sense of distributions \cite{IMG}. The trace
can be explicitly calculated as%
\begin{equation}
tr\left[  T_{\chi}(g)\right]  =\dfrac{\cos(\eta\rho+n\theta)}{2\left\vert
\sinh(\eta+i\theta)\right\vert ^{2}}. \label{eq.trsl(2,C)}%
\end{equation}
Therefore, the expression for the delta on $SL(2,C)$ explicitly is%
\begin{equation}
\delta(g)=\frac{1}{8\pi^{4}}\sum_{n}\int d\rho(n^{2}+\rho^{2})\dfrac{\cos
(\rho\eta+n\theta)}{\left\vert \sinh(\eta+i\theta)\right\vert ^{2}}.
\label{deltaExplicit}%
\end{equation}
Let us consider the integrand in equation (\ref{invFour}). Using equation
(\ref{Four}) in it we have
\begin{align}
Tr(F(\chi)T_{\chi}(g^{-1}))\chi\bar{\chi}  &  =\chi\bar{\chi}\int f(\acute
{g})Tr(T_{\chi}(\acute{g})T_{\chi}(g^{-1}))d\acute{g}\nonumber\\
&  =\chi\bar{\chi}\int f(\acute{g})Tr(T_{\chi}(\acute{g}g^{-1}))d\acute{g}.
\label{SignChi}%
\end{align}
But, since the trace is insensitive to an overall sign of $\chi$, so are the
terms of the Fourier expansion of the $L^{2}$ functions on $SL(2,C)\ $and
$CS^{3}$.

\section{Unitary Representations of $SO(4,C)$}

The group $SO(4,C)$ is related to its universal covering group $SL(2,C)\times
SL(2,C)$ by the relationship $SO(4,C)\approx\frac{SL(2,C)\times SL(2,C)}%
{Z^{2}}$. The map from $SO(4,C)$ to $SL(2,C)\times SL(2,C)$ is given by the
isomorphism between complex four vectors and $GL(2,C)$ matrices. If
$X=(a,b,c,d)$ then $G:C^{4}\longrightarrow GL(2,C)\ $can be defined by the
following:%
\[
G(X)=\left[
\begin{array}
[c]{cc}%
a+ib & c+id\\
-c+id & a-ib
\end{array}
\right]  .
\]
It can be easily inferred that $\det G(X)=a^{2}+$ $b^{2}+c^{2}+d^{2}$ is the
Euclidean norm of the vector $X$. Then, in general a $SO(4,C)$ rotation of a
vector $X$ to another vector $Y$ is given in terms of two arbitrary
$SL(2,C)\ $matrices $g_{L~B}^{~~A},~g_{R~B^{^{\prime}}}^{~~A^{^{\prime}}}\in
SL(2,C)$ by
\[
G(Y)^{AA^{^{\prime}}}=g_{L~B}^{~~A}g_{R~B^{^{\prime}}}^{~A^{^{\prime}}}%
G^{AB}(X),
\]
where $G^{AB}(X)$ is the matrix elements of $G(X)$. The above transformation
does not differentiate between $(L_{B}^{A},R_{B^{^{\prime}}}^{A^{^{\prime}}})$
and $(-L_{B}^{A},-R_{B^{^{\prime}}}^{A^{^{\prime}}})$ which is responsible for
the factor $Z_{2}$ in $SO(4,C)\approx\frac{SL(2,C)\times SL(2,C)}{Z^{2}}$.

The unitary representation theory of the group $SL(2,C)\times SL(2,C)$ is
easily obtained by taking the tensor products of two Gelfand-Naimarck
representations of $SL(2,C)$. The Fourier expansion for any function
$f(g_{L},g_{R})$ of the universal cover is given by%
\[
f(g_{L},g_{R})=\frac{1}{64\pi^{8}}\int\chi_{L}\bar{\chi}_{L}\chi_{R}\bar{\chi
}_{R}F(\chi_{L},\chi_{R})T_{\chi}(g_{L}^{-1})T_{\chi}(g_{R}^{-1})d\chi
_{L}d\chi_{R},
\]
where $\chi_{L}=\frac{n_{L}+i\rho_{L}}{2}$ and $\chi_{R}=\frac{n_{R}+i\rho
_{R}}{2}$. The Fourier expansion on $SO(4,C)$ is given by reducing the above
expansion such that $f(g_{L},g_{R})=f(-g_{L},-g_{R})$. From equation
(\ref{eq.trsl(2,C)}) I have%
\[
tr\left[  T_{\chi}(-g)\right]  =(-1)^{n}tr\left[  T_{\chi}(-g)\right]  ,
\]
where $\chi=\frac{n+i\rho}{2}$. Therefore%
\[
f(-g_{L},-g_{R})=\frac{1}{8\pi^{4}}\int\chi_{L}\bar{\chi}_{L}\chi_{R}\bar
{\chi}_{R}F(\chi_{L},\chi_{R})(-1)^{n_{L}+n_{R}}T_{\chi}(g_{L}^{-1})T_{\chi
}(g_{R}^{-1})d\chi_{L}d\chi_{R}.
\]
This implies that for $f(g_{L},g_{R})=f(-g_{L},-g_{R}),$ I must
have$~(-1)^{n_{L}+n_{R}}$ $=1$. From this, I can infer that the representation
theory of $SO(4,C)$ is deduced from the representation theory of
$SL(2,C)\times SL(2,C)$ by restricting $n_{L}+n_{R}$ to be even integers. This
means that $n_{L}$ and $n_{R}$ should be either both odd numbers or even
numbers. I would like to denote the pair $(\chi_{L},\chi_{R})$ ($n_{L}+n_{R}$
even) by $\omega$.

There are two Casimir operators available for $SO(4,C),$ namely $\varepsilon
_{IJKL}\hat{B}^{IJ}\hat{B}^{KL}$ and $\eta_{IK}\eta_{JL}\hat{B}^{IJ}\hat
{B}^{KL}$. The elements of the representation space $D_{\chi_{L}}\otimes$
$D_{\chi_{R}}$ are the eigen states of the Casimirs. On them, the operators
reduce to the following:
\begin{equation}
\varepsilon_{IJKL}\hat{B}^{IJ}\hat{B}^{KL}=\frac{\chi_{L}^{2}-\chi_{R}^{2}}%
{2}~~\text{and}%
\end{equation}%
\begin{equation}
\eta_{IK}\eta_{JL}\hat{B}^{IJ}\hat{B}^{KL}=\frac{\chi_{L}^{2}+\chi_{R}^{2}%
-2}{2}.
\end{equation}

\section{Unitary Representations of $SU(1,1)$}

The unitary representations of $SU(1,1)\approx SL(2,R)$, given in
Ref:\cite{NJaVAUK}, is defined similar to that of $SL(2,C)$. The main
difference is that the $D_{\chi}$ are now functions $\phi(z)$ on $C^{1}$. The
representations are indicated by a pair $\chi=$ $(\tau,\varepsilon),$
$\varepsilon$ is the parity of the functions ($\varepsilon$ is $0$ for even
functions and $\frac{1}{2}$ for odd functions) and $\tau$ is a complex number
defining the homogeneity:%
\[
\phi(az)=\left\vert a\right\vert ^{2\tau}sgn(a)^{2\varepsilon}\phi(z),
\]
where $a$ is a real number. Because of homogeneity the $D_{\chi}$ functions
can be related to the infinitely differentiable functions $\phi(e^{i\theta})$
on $S^{1}$ where $\theta$ is the coordinate on $S^{1}$. The representations
are defined by%
\begin{equation}
T_{\chi}(g)\phi(e^{i\theta})=(\beta e^{i\theta}+\bar{\alpha})^{\tau
+\varepsilon}(\bar{\beta}e^{-i\theta}+\alpha)^{\tau-\varepsilon}\phi
(\frac{\alpha z+\bar{\beta}}{\beta z+\bar{\alpha}}). \label{T-SL(2,R)}%
\end{equation}
There are two types of the unitary representations that are relevant for
quantum general relativity: the continuous series and the discrete series. For
the continuous series $\chi=(i\rho-\frac{1}{2},\varepsilon),$ where $\rho$ is
a non-zero real number. Let me denote the continuous series representations
with suffix or prefix $c,$ for example $T_{\chi}^{c}$.

There are two types of discrete series representations which are indicated by
signs $\pm.$ They have their respective homogeneity as $\chi_{\pm
}=(l,\varepsilon_{l}^{\pm})$ where $\varepsilon_{l}^{\pm}=\pm1$ is defined by
the condition $l\pm\varepsilon_{l}^{\pm}$ is an integer. Let me denote the
representations as $T_{l}^{+}$ and $T_{l}^{-}$. The $T_{l}^{+}$ ($T_{l}^{-}$)
representations can be re-expressed as linear operators on the functions
$\phi_{+}(z)$($\phi_{-}(z)$) on $C^{1}$ that are analytical inside (outside)
the unit circle. The $T_{l}^{\pm}(g)$ are defined as%
\[
T_{l}^{\pm}(g)\phi_{\pm}(z)=\left\vert \beta z+\bar{\alpha}\right\vert
^{2l}\phi_{\pm}(\frac{\alpha z+\bar{\beta}}{\beta z+\bar{\alpha}}).
\]
The inner products are defined by%
\begin{align*}
(f_{1},f_{2})_{c}  &  =\frac{1}{2\pi}\int_{0}^{2\pi}f_{1}(e^{i\theta
})\overline{f_{2}(e^{i\theta})}d\theta,\\
(f_{1},f_{2})_{+}^{l}  &  =\frac{1}{\Gamma(-2l-1)}\iint\limits_{\left\vert
z\right\vert <1}(1-\left\vert z\right\vert )^{-2l-2}f_{1}(z)f_{2}%
(z)\frac{dzd\bar{z}}{2\pi i},\\
(f_{1},f_{2})_{-}^{l}  &  =\frac{1}{\Gamma(-2l-1)}\iint\limits_{\left\vert
z\right\vert >1}(1-\left\vert z\right\vert )^{-2l-2}f_{1}(z)f_{2}%
(z)\frac{dzd\bar{z}}{2\pi i}.
\end{align*}
The Fourier transforms are defined for the unitary representations by%
\begin{align*}
F_{c}(\chi)  &  =\int f(g)T_{\chi}^{c}(g)dg,\\
F_{+}(l)  &  =\int f(g)T_{l}^{+}(g)dg,~~\text{and}\\
F_{-}(l)  &  =\int f(g)T_{l}^{-}(g)dg,
\end{align*}
where $dg$ is the bi-invariant measure on the group.

The inverse Fourier transform is defined by
\[
f(g)=\frac{1}{4\pi^{2}}\left\{
\begin{array}
[c]{c}%
\sum_{l\in\frac{1}{2}N_{0}}(l+\frac{1}{2})Tr[F_{+}^{\dagger}(l)(T_{l}%
^{+}(g))+F_{-}^{\dagger}(l)(T_{l}^{-}(g))]\\
+\sum_{\varepsilon}\int_{0}^{\infty}\rho Tr[F(\chi)T_{\rho}^{\dagger}]\tanh
\pi(\rho+i\varepsilon)d\rho
\end{array}
\right\}  .
\]
The $T_{(\tau,\varepsilon)}$ is equivalent to $T_{(-\tau-1,\varepsilon)}$. The
Casimir operator for the $T_{\chi}$ representations (all) can be defined
similar to $SU(2)$ and its eigen values are%
\[
C=\tau(\tau+1),
\]
where the $\tau$ comes from $\chi=(\tau,\varepsilon).$ The $\tau$ in this
section is related to the $\chi$ in the representations of $SL(2,C)\ $by
$\chi=\tau+\frac{1}{2}$. The expressions for the Casimirs of the two groups
differ by a factor of $4$.

\end{document}